\documentclass[fleqn,usenatbib]{mnras}

\usepackage{newtxtext,newtxmath}
\usepackage[T1]{fontenc}
\usepackage[dvipsnames]{xcolor} 
\DeclareRobustCommand{\VAN}[3]{#2}
\let\VANthebibliography\thebibliography
\def\thebibliography{\DeclareRobustCommand{\VAN}[3]{##3}\VANthebibliography}

\usepackage{graphicx}	% Including figure files
\usepackage{amsmath}	% Advanced maths commands
\usepackage{pdflscape}
\usepackage{comment}
\usepackage{subcaption}  % Put this in your preamble
\usepackage{threeparttable}

\newcommand\boldred[1]{\textcolor{red}{}}

\title[AGE-PRO models]{The ALMA Survey of Gas Evolution of PROtoplanetary Disks (AGE-PRO): Constraints on disk turbulence, fragmentation velocity, and inner pebble fluxes}

\author[Lilian Luo et al.]{
Lilian Luo,$^{1}$\thanks{E-mail: lilian.luo.22@ucl.ac.uk},
Paola Pinilla$^{1}$,
Camila Pulgarés$^{2}$,
Laura M. Pérez$^{2}$,
Miguel Vioque$^{3}$,
Nicolás T. Kurtovic$^{5,6}$,
\newauthor
Anibal Sierra$^{1}$,
Carolina Agurto-Gangas$^{7}$,
Rossella Anania $^{8}$,
John Carpenter$^{4}$,
Lucas A. Cieza$^{9,12}$,
\newauthor
Dingshan Deng$^{10}$,
James Miley$^{11,12,13,4,16}$,
Ilaria Pascucci$^{10}$,
Giovanni P. Rosotti$^{8}$,
Benoît Tabone $^{14}$,
Ke Zhang $^{15}$
\\
$^{1}$Mullard Space Science Laboratory, University College London, Holmbury St Mary, Dorking, Surrey RH5 6NT, UK\\
$^{2}$Departamento de Astronomía, Universidad de Chile, Camino El Observatorio 1515, Las Condes, Santiago, Chile\\
$^{3}$European Southern Observatory, Karl-Schwarzschild-Str. 2, 85748 Garching bei München, Germany \\
$^{4}$Joint ALMA Observatory, Alonso de Córdova 3107, Vitacura, Santiago 763-0355, Chile \\
$^{5}$ Max Planck Institute for Extraterrestrial Physics, Giessenbachstrasse 1, D-85748 Garching, Germany \\
$^{6}$ Max-Planck-Institut fur Astronomie (MPIA), Konigstuhl 17, 69117 Heidelberg, Germany \\
$^{7}$Departamento de Física, Universidad Técnica Federico Santa María, Vicuña Mackenna 3939, San Joaquín, Santiago de Chile, Chile \\
$^{8}$Dipartimento di Fisica, Università degli Studi di Milano, Via Celoria 16, I-20133 Milano, Italy \\
$^{9}$Instituto de Estudios Astrofísicos, Universidad Diego Portales, Av. Ejercito 441, Santiago, Chile \\
$^{10}$Lunar and Planetary Laboratory, the University of Arizona, Tucson, AZ 85721, USA \\
$^{11}$Departamento de Física, Universidad de Santiago de Chile, Avenida Victor Jara 3659, Santiago, Chile \\
$^{12}$Millennium Nucleus on Young Exoplanets and their Moons (YEMS), Chile \\
$^{13}$ Center for Interdisciplinary Research in Astrophysics and Space Exploration (CIRAS), Universidad de Santiago, Chile \\
$^{14}$Université Paris-Saclay, CNRS, Institut d’Astrophysique Spatiale, 91405 Orsay, France \\
$^{15}$Department of Astronomy, University of Wisconsin-Madison, 475 N Charter St, Madison, WI 53706, USA \\
$^{16}$European Southern Observatory, Alonso de Córdova 3107, Vitacura, Santiago, Chile
}

\date{Accepted by MNRAS, February 2026}

\pubyear{2026}

\begin{document}
\label{firstpage}
\pagerange{\pageref{firstpage}--\pageref{lastpage}}
\maketitle

\begin{abstract}
How substructures and disk properties affect dust evolution and the delivery of solids and volatiles into planet-forming regions remains an open question.
We present results from tailored dust evolution modeling of the AGE-PRO ALMA large program, a sample of 30 protoplanetary disks spanning different evolutionary stages. Visibility fitting of the AGE-PRO ALMA data (at 1.3\,mm) reveals that approximately half of the disks exhibit radial substructures. Combined with stellar properties, disk inclinations, and gas mass estimates from CO isotopologues and N$_2$H$^+$, this well-characterized set of disks provides an ideal testbed to constrain dust evolution models across different ages and disk morphologies. Using the dust evolution code \texttt{DustPy}, we simulate dust evolution in each disk under four model configurations, varying two key free parameters: the turbulent viscosity ($\alpha = 10^{-4}, 10^{-3}$) and fragmentation velocity ($v_{\rm{frag}} = 1 \mathrm{m\,s^{-1}}, 10 \mathrm{m\,s^{-1}}$). Pressure traps are incorporated by perturbing the gas surface density based on the continuum intensity profiles, and synthetic observations generated with \texttt{RADMC-3D} are compared to these profiles.
While no single model fits all disks, nearly half are best reproduced by the configuration with low turbulence and low fragmentation velocity ($\alpha = 10^{-4}, v_{\rm{frag}} = 1\,\mathrm{m\,s^{-1}}$). Models of smooth disks underpredict dust mass, possibly indicating unresolved substructures. Pebble fluxes into inner disk regions correlate more strongly with disk age than with the presence of substructures, highlighting time-dependent dust transport as a key factor in shaping inner disk composition.
Our results also provide a comparative baseline for interpreting multiwavelength and JWST water vapor observations.

\end{abstract}


\section{Introduction}

Dust in protoplanetary disks is crucial to our understanding of planet formation, both as the raw material for solids and as a tracer of disk structure and evolution. Its growth, fragmentation, and radial transport influence the formation of planetesimals and play an important role in the delivery of volatiles into the inner few au of the disk, where terrestrial planets are thought to form. The initial properties of dust grains — inherited from the interstellar medium — evolve as grains collide, stick, bounce, fragment, and drift within the gas disk \citep[e.g.,][]{whipple1972, 1977Weidenschilling, 2005Dominik&Dullemond, Testi_2014, 2012Birnstiel, pinilla2012, Pinilla_2020, Birnstiel_2024}. These processes are strongly influenced by the local disk environment, particularly the level of turbulence and presence of pressure structures \citep[e.g.,][]{pinilla2012, Andrews_2018, 2020Flaherty}. Understanding these processes is crucial for constraining the initial conditions of planet formation.

Recent high-resolution ALMA observations have revealed a large diversity in disk morphologies. While some disks appear compact and smooth in their dust emission \citep[e.g.,][]{long2019, Guerra_Alvarado_2025}, others are large and exhibit concentric rings, gaps, asymmetries, or azimuthal structures \citep[e.g.,][]{Andrews_2018, Andrews2020, 2018Long, 2018Huang}. These features are often interpreted as signatures of local pressure maxima, which can trap dust grains and slow their inward drift \citep[][]{whipple1972, pinilla2012}, allowing solids to grow to larger sizes and remain trapped in the outer disk for longer periods. However, the degree to which these substructures influence overall dust retention and growth efficiencies, as well as the formation of terrestrial planets in the innermost parts of disks, remains an open question. 

From a theoretical standpoint, different key parameters control grain growth: (1) the coupling of dust particles to the gas, which is quantified by the Stokes number \citep{2012Birnstiel}, defined in the disk midplane as:
\begin{equation}
\mathrm{St} = \frac{a \, \rho_{\mathrm{s}}}{\Sigma_{\mathrm{g}}} \frac{\pi}{2},
\label{eq:stokes}
\end{equation}

\noindent where \(a\) is the grain size, \(\rho_{\mathrm{s}}\) is the volume density of the dust grains, and \(\Sigma_{\mathrm{g}}\) is the gas surface density. (2) The turbulent viscosity parameter $\alpha$, which sets the turbulent relative velocities and dust diffusion. (3) The fragmentation velocity $v_{\rm{frag}}$, which sets the threshold above which particles are destroyed upon collision. Both parameters, $\alpha$ and $v_{\rm{frag}}$, are difficult to constrain from observations. The $\alpha$ parameter \citep{shakura1973black} is typically assumed to range from $10^{-2}$ in active accretion regions to $10^{-4}$ or lower in dead zones \citep{flock2013, 2020Flaherty}. Laboratory and numerical studies suggest $v_{\rm{frag}}$ values in the range of $1$–$10$\,m\,s$^{-1}$, depending on material composition and temperature \citep[e.g.,][]{2008BlumWurm, Wada2009, 2019MusiolikWurm}. We discuss this further in Sect. \ref{sect:methods}.

In this work, we focus on turbulence values ($\alpha$ = $10^{-3}-10^{-4}$), consistent with recent observational upper limits  \citep[e.g.,][]{Villenave_2022, Villenave2025}, while higher $\alpha$ values ($\sim 10^{-2}$) are unlikely given constraints from vertical settling.

Interpreting the interplay of dust growth, fragmentation, and transport processes in disks requires large, uniform samples with well-characterized gas and dust properties. A key requirement is also a sample that spans a range of stellar ages and environments, allowing us to disentangle the effects of disk evolution over time from initial conditions. The ALMA Survey of Gas Evolution of PROtoplanetary Disks (AGE-PRO, PI: K. Zhang)
provides an unprecedented opportunity to test dust evolution models in such a context. It targets 30 Class I–II protoplanetary disks from three nearby star-forming regions — Ophiuchus, Lupus, and Upper Sco — which together span a broad range of ages (from $\sim$0.5 to 5 Myr). The AGE-PRO sample is designed such that the stellar mass range is narrow ($\sim0.3 - 0.8 M_{\odot}$), to focus on the disk evolution and to avoid dependences on stellar properties. All disks were observed uniformly in 1.3\,mm continuum and molecular line tracers.

Except for the disks in Ophiuchus, gas mass estimates for each disk are derived from CO isotopologues and N$_{2}$H$^{+}$ emission, accounting for CO depletion and chemical evolution effects as presented in \citep{Trapman_2022}.
For the Ophiuchus disks, a combination of  $^{12}$CO,  $^{13}$CO, C$^{17}$O, and C$^{18}$O was used. This is because, for young disks, a combination of isotopologues can be used, as freeze-out and isotopologue-selective dissociation are less important \citep[][]{zhang2020, zhang2025, Trapman_2025_gas}. The resulting estimates are consistent with those inferred from C$^{18}$O using self-consistent disk models \citep{2025Deng}.

Radial intensity profiles of continuum emission were extracted directly from modeling the visibilities as part of a uniform modeling effort by the AGE-PRO collaboration \citep{2025_Vioque}, using the \texttt{frankenstein} tool \citep[hereafter FRANK,][]{2020Jennings}.  These fits allow us to probe spatial structures on scales even finer than that of the synthesized beam achieved in imaging, and reveal that approximately half the sample exhibits a deviation from a smooth profile, while the remaining disks were consistent with being smooth at the AGE-PRO resolution ($\sim0.2-0.5^{\prime \prime}$). This diversity - in structure, gas mass, and age - makes AGE-PRO a uniquely powerful laboratory for constraining dust evolution physics. In particular, the well-constrained gas masses allow us to more accurately estimate the Stokes number of observed grains (see Eq.~\ref{eq:stokes}), enabling a direct link between local disk conditions and the aerodynamic evolution of solids. 

In this study, we test whether standard dust evolution models can reproduce the diversity of dust and disk properties observed in AGE-PRO. Specifically, we investigate whether a single set of $\alpha$ and $v_{\rm{frag}}$ parameters can simultaneously explain the observed trends in dust mass, dust disk radius $(R_{90\%})$, and spectral index ($\alpha_{\rm{mm}}$). Using the \texttt{DustPy} code \citep{Stammler_2022}, we simulate the dust evolution for each disk under four physical configurations, varying $\alpha=[10^{-4}, 10^{-3}]$ and $v_{\rm frag} = [1,\,10]\,\mathrm{m\,s^{-1}}$. For disks identified as substructured from the visibility modeling, we incorporate pressure bumps into the gas surface density based on the FRANK-derived profiles. These structures act as localized pressure traps and are modeled as Gaussian perturbations superposed on a smooth profile \citep{Pinilla_2020}.
To compare models to observations, we generate synthetic ALMA images using the RADMC-3D radiative transfer code \citep[][]{dullemond2012radmc}, using the FRANK minimum resolvable scale. This approach ensures consistent observational diagnostics across both data and models. We then compare the resulting radial intensity profile, dust mass, dust disk sizes, and spectral index for each configuration, and assess which models best reproduce the observed properties.

\begin{figure*}
    \centering
    \includegraphics[width=0.96\textwidth]{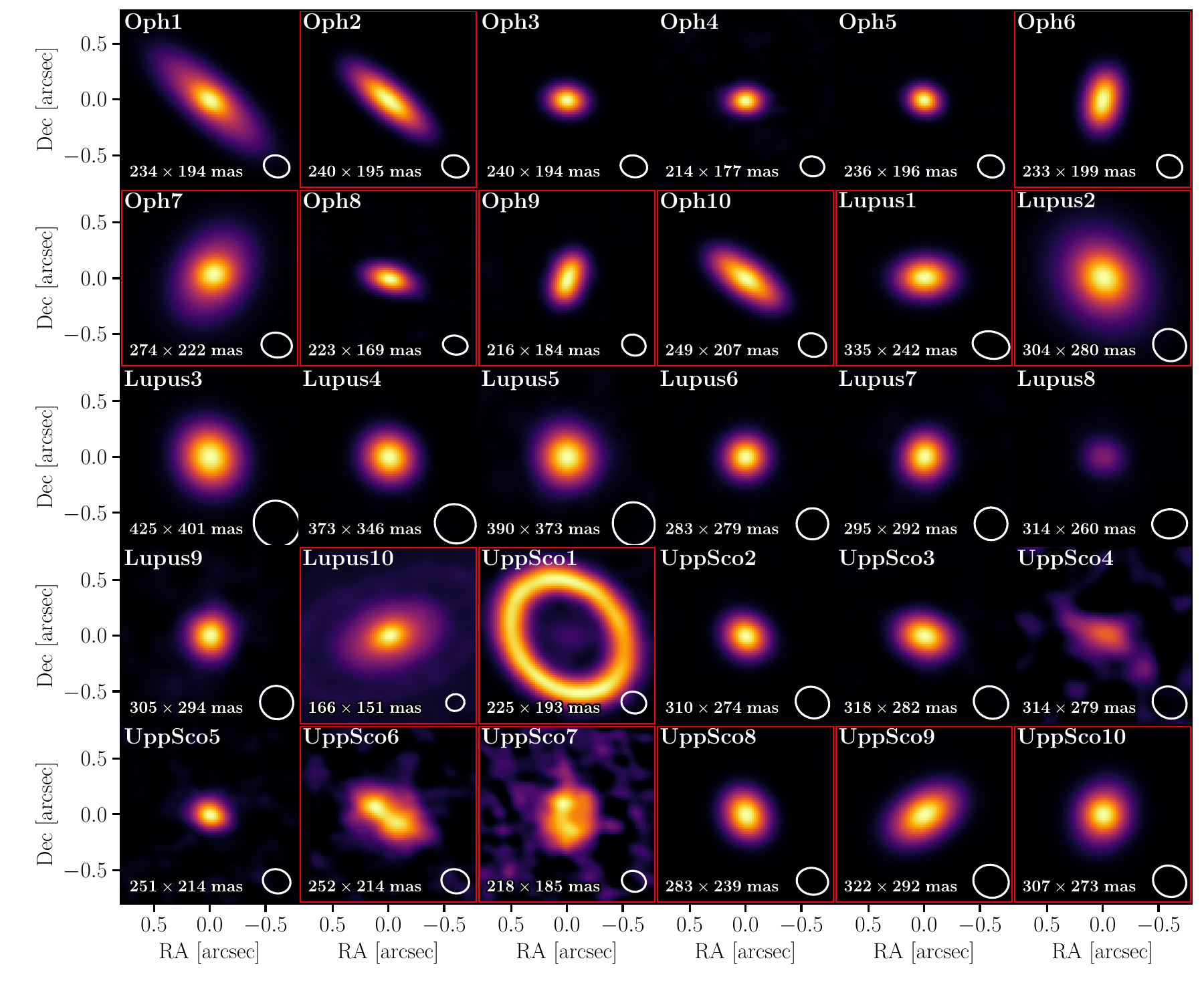}
    \caption{
    Dust continuum emission of all the disks in the AGE-PRO sample, in ALMA Band 6 (1.3\,mm). The disks with substructures inferred from the visibility models are enclosed by a red square. The beam size and shape is shown in each panel.The disk ID follows the definition in \protect\cite{2025Ruiz_Rodriguez,2025Deng,2025Agurto-Gangas}, for Ophiuchus, Lupus, and Upper Sco, respectively.}
    \label{fig:galleryAGEPRO}
\end{figure*}

Beyond testing which models best match each individual disk, we aim to explore broader trends across the sample. We investigate whether low turbulence and fragmentation velocities are systematically favored in reproducing observed dust disk properties. We assess whether substructures are required to explain the observed retention of dust mass, or if smooth models can reproduce these features without invoking trapping. In doing so, we build on the findings of \citet{Kurtovic2025}, who showed that the AGE-PRO sample is consistent with simulations that have either weak or strong dust traps and that the evolution of spectral index in the sample is also suggestive of an unresolved population of dust traps. Finally, we analyze the radial distribution of pebble fluxes to estimate how much solid material is transported into the inner few au—a quantity relevant for interpreting recent JWST detections of cool water excess in inner disks \citep[e.g.,][]{banzatti2023,2023SierraGrant,gasman2024}.

We also note that similar qualitative conclusions have recently been reached by \citet{Tong_2025}, who used DustPy models to explore the coupled role of turbulence and fragility of dust in a different disk sample. Their findings favor low turbulence and fragile dust to best reproduce multiwavelength millimeter observations.

This paper is organized as follows. In Sect.~\ref{sect:obs_sample}, we describe our observational sample. In Sect.~\ref{sect:methods}, we describe our disk models, the \texttt{DustPy} setup, and our assumptions for gas and dust evolution, including treatment of substructures and radiative transfer. In Sect.~\ref{sect:results}, we present the results of our simulations and comparisons with observed dust masses, sizes, and spectral indices. In Sect.~\ref{sect:discussion}, we discuss the implications of our results for dust retention, substructure interpretation, and pebble delivery. Sect.~\ref{sect:conclusion} summarizes our conclusions and outlines future work.

\section{AGE-PRO as a laboratory of dust evolution
models}  \label{sect:obs_sample}

We model the disks of the AGE-PRO ALMA large program, which aims to systematically trace the evolution of gas disk mass and size throughout the lifetime of protoplanetary disks \citep{zhang2025}. Gas plays a dominant role in regulating the dynamics, distribution and growth of dust. Gas interacts with dust primarily through aerodynamic drag, affecting the radial drift and vertical settling of dust. An accurate gas mass and disk size, and by extension gas surface density, is key when modeling these processes, as they will determine where the dust concentrates and grows. Therefore, the mass of protoplanetary disks is one of the most critical properties influencing their evolution into potential planetary systems \citep{Birnstiel_2024}. 

In addition to the first data release of AGE-PRO, follow-up observations with ALMA Bands 3 and 5 were obtained (Pulgarés et al., in prep.), providing a rich dataset for multi-wavelength analysis and additional constraints on dust growth across evolutionary stages.We incorporate these multi-wavelength data into our analysis of spectral indices and dust disk sizes (see Sect. \ref{sect:results}). Additional JWST observations of the disks allow us to test how the water abundance into the terrestrial planet-forming region is influenced by pebble fluxes to the inner disk and substructures \citep{Kalyaan_2021, kalyaan2023, Mah_2024}.

\cite{2025_Vioque} performed visibility fitting of the Band 6 (1.3\,mm) observations from the AGE-PRO  \citep[][]{2025Deng, 2025Agurto-Gangas, 2025Ruiz_Rodriguez}, retrieving disk geometries, dust-disk radii, and azimuthally symmetric FRANK radial profiles of the intensity of the dust continuum emission. The models with FRANK provide non-parametric fits assuming azimuthal symmetry of the visibility data, enabling the recovery of sub-beam features (compared to \texttt{CLEAN} image) while leveraging the full sensitivity of the data \citep[e.g.,][]{sierra2024}. This allowed for a more detailed view of the potential substructures in disks.  \cite{2025_Vioque} finds that roughly half of the 30 disks in AGE-PRO show substructure at our observational resolution, allowing us to study the effects of rings and gaps in our models. Figure~\ref{fig:galleryAGEPRO} shows the dust continuum emission of all the disks in the AGE-PRO sample, and highlights (with a red square) the disks where substructures were detected.

\section{Methods} \label{sect:methods}

\subsection{Dust Evolution Models with DustPy}
In this work, we model the AGE-PRO disks using the dust evolution code \texttt{DustPy} \citep{Stammler_2022}. For given stellar and disk parameters, \texttt{DustPy} tracks the evolution of gas and dust surface densities in disks, by solving the equations for viscous evolution of gas, advection and diffusion of the dust. \texttt{DustPy} models dust growth by solving the Smoluchowski equation \citep{Stammler_2022}, which captures the collisional growth of particles in 1D. It models the radial evolution of dust and gas while assuming vertical hydrostatic equilibrium. In our models, we only evolve the dust and assume a fixed gas surface density. This ensures that the assumed gas masses reflect the AGE-PRO mass estimates. We further motivate this choice by the AGE-PRO results, which suggest the dust appears to evolve and disperse faster than the gas.

In addition, based on the gas evolution models from the AGE-PRO collaboration \citep[]{Tabone_2025, 2025Anania}, it is still unclear which mechanism is dominating the gas evolution. There is growing evidence showing that viscous evolution alone cannot account for the observed spread in the $M_{\mathrm{acc}}$–$M_{\mathrm{disk}}$ relation, pointing to a potentially significant role for disk winds \citep[e.g.,][]{Tabone_2025}. The choice of keeping the gas surface density constant allows us to isolate the impact of dust evolution physics and substructure morphology, but does not account for potential feedback between dust and gas evolution or external effects such as photo-evaporation.

We model the gas surface density profile using the conventionally adopted description of a viscous accreting disk \citep{LyndenBell}. This profile is described as a power law with an exponential taper:

\begin{equation} 
\Sigma_{\text{gas}}(r) = \Sigma_0 \left(\frac{r}{R_c}\right)^{-\gamma} \exp\left(-\left(\frac{r}{R_c}\right)\right),
\label{eq:gas_surface_density}
\end{equation}

\noindent where $\Sigma_0$ is the value of the surface density at the characteristic radius $R_c$, which is the location where the exponential taper begins to dominate, and $\gamma$ is the power-law exponent that determines the slope of the surface density profile at  radii lower than $R_c$. For each disk,  $\Sigma_0$ is tailored to match the gas disk mass as obtained in \cite{Trapman_2025_gas}. In practice, we adopted the central reported value of the gas mass, and note that associated uncertainties are large and not propagated in our modeling.

\begin{figure*}
    \centering
    \includegraphics[width=0.48\textwidth]{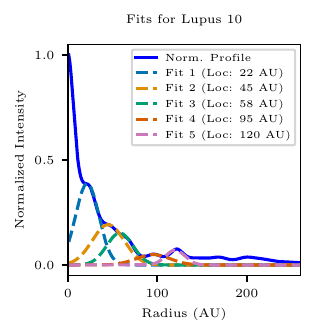}
    \hfill
    \includegraphics[width=0.48\textwidth]{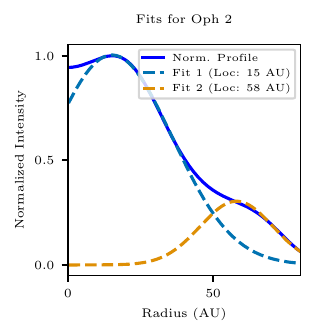}
    \caption{Two examples of Gaussian fits applied to the normalized radial brightness profiles obtained from FRANK using \texttt{curve\_fit}. The left panel shows Lupus 10, an extended disk which exhibits multiple prominent rings and gaps, while the right panel shows Oph 2, featuring two distinct bumps. These fits are used to infer the location and relative amplitude of pressure bumps for the dust evolution models.}
    \label{fig:gaussianfits}
\end{figure*}

To replicate the substructures observed in the dust continuum, we create dust traps by introducing perturbations in the gas surface density that can act as gaps or rings \citet{Pinilla_2020}. These pressure bumps are described by: 

\begin{equation}
B(r) = A \exp\left(-\frac{(r - r_p)^2}{2w^2}\right),
\end{equation}

\noindent where $A$ is the amplitude, $r_p$ is the radial centre of the pressure bump, $w$ is the width of the perturbation, which is a factor of the gas scale height. We tailor the depth and width of the gap(s) and/or ring(s) based on how large the substructures appear in the intensity profile. Depending on the complexity of the substructures, a number of rings and gaps ($n_s$) are added to the unperturbed density profile, such as

\begin{equation}
\Sigma'_{\text{gas}}(r) = \Sigma_{\text{gas}}(r) \left[1 \pm \sum_{n_s} B_{n_s}\right],
\end{equation}

\noindent where the sign determines if it is a gap or a ring added to the perturbation.

\subsection{Using continuum profile as prior for dust trap strength}

Although \cite{2025_Vioque} provided the radial intensity profile, the relationship between the dust continuum intensity and the dust surface density is not direct, and it is even more indirect with the gas surface density. The brightness profile and dust surface density relate through the Planck function as:

\begin{equation}
    I(r) = B_{v} (T(r))[1-e^{-\tau_{v}(r)}],
\end{equation}
\noindent where the disk temperature ($T(r)$) is obtained from the radiative transfer models described below. $I(r)$ is related to the dust surface density through $\tau_{v}(r)$, the optical depth for a given frequency at a radial position $r$. As the opacity is dominated by dust, the  optical depth is given by: 
\begin{equation}
    \tau_{v} = \sigma(r,a)k_{v}/\cos(i),
\end{equation}
\noindent where $\sigma(r,a)$ is the dust surface density at $r$ and dependent on particle size $a$, and $k_{v}$ is the dust opacity for each grain size simulated in the dust evolution models, and $cos(i)$ accounts for the inclination angle $i$ of the disk.

To model the observed substructures in the disks, we use the Levenberg–Marquardt algorithm \citep{Gavin2019}, a method for non-linear least squares fitting, to fit Gaussians to the rings in the normalized brightness profile from AGE-PRO \citep{2025_Vioque}. The relative amplitude of these structures was used to scale the perturbations in the gas surface density. We use the function \texttt{curve$_{-}$fit} \citep{more1980user} from \texttt{scipy.optimize} and use visual inspection of the profiles to find the initial guesses ($r_{p0}$) for the centre of ring locations. 

To fit the Gaussian function to the normalized brightness profile $(x_i, y_i)$, the parameters $A$ (amplitude), $r_p$ (radial location of the centre of the ring), and $w$ (width of the Gaussian) are optimized by minimizing the sum of squared residuals, $S(A, r_p, w)$, given by:
\begin{equation}
    S(A, r_p, w) = \sum_{i=1}^{N} \left[ y_i - f(x_i; A, r_p, w) \right]^2,
\end{equation}

After we obtain the best-fit ring locations of our radial profile, we scale the amplitude of the pressure bumps in the perturbed gas surface density based on their relative amplitude. We only selected the bumps that are at least 10\% of the peak. Figure~\ref{fig:gaussianfits} shows two examples, Lupus 10 and Oph 2, where Gaussian fits are applied to the normalized radial brightness profiles obtained from FRANK. The fits capture the location and relative contrast of rings and gaps in the disk, which are then used to place pressure bumps in the gas surface density for the dust evolution models. A summary of the fitted $r_p$ is provided in Table~\ref{tab:substructures}.

\subsection{DustPy Parameters and Assumptions}
The main parameters used in the \texttt{DustPy} simulations are summarized in Table~\ref{tab:DustPy_params}, including the initial grain size distribution and the free parameters turbulence parameter $\alpha$, and fragmentation velocity.

For most disks in our sample, $R_c$ values are based on the values reported by \cite{2025_Trapman_size}. However, for the more massive disks Oph 1 and Oph 2, we increased the $R_c$ values relative to those reported in that work. This adjustment was necessary to prevent unphysical artifacts in the synthetic continuum images, likely caused by the very long drift timescales at the disk outer edge. Using smaller $R_c$ values led to overly sharp outer profiles and non-physical emission features.

The choice of $r_{\text{max}}$ varies across the sample and is motivated by the extent of the observed continuum emission in each disk and how massive they are. Large disks with larger radial emission are assigned larger computational domains (600 or 1000 AU), ensuring that the outer disk structure is fully captured and effects at the edge of the radial boundary are avoided. This is done for disks Oph 1, 2, Lupus 10. The rest of the disks are modeled with $r_{\text{max}} = 300$ AU, which provides sufficient coverage of the dust evolution region while optimizing computational efficiency.

In Table ~\ref{tab:substructures} we summarize the target disk properties, such as stellar mass, disk age, and gas mass, which are used as fixed inputs in the simulations.

\begin{table}
    \centering
    \caption{Parameter Space for \texttt{DustPy} Simulations with the varied parameters highlighted.}
    \label{tab:DustPy_params}
    \begin{tabular}{|l|l|}
        \hline
        \textbf{Parameter} & \textbf{Value} \\
        \hline
        Radial grid ($r_{\text{min}}$) & 1 AU \\
        Radial grid ($r_{\text{max}}$) & 300/600/1000 AU \\
        Number of radial grid points (Nr) & 100/200/300 (scaled with $r_{\text{max}}$) \\
        Initial dust size distribution & MRN Power-law \citep{MRN} \\
        q & 3.5 \\
        Amin, Amax & $10^{-5}$, 10 cm \\
        \textbf{Fragmentation velocity} ($v_{\mathrm{frag}}$) & $1$, $10\,\mathrm{m\,s^{-1}}$ \\
        \textbf{Turbulence parameter} ($\alpha$) & $10^{-3}$, $10^{-4}$ \\
        Disk age & 0.5, 1.5, 5 Myr \\
        Gas masses & From \cite{Trapman_2025_gas} \\
        Initial dust mass & 1\% of gas mass \\
        Critical radius ($R_c$)  & Based on \cite{2025_Trapman_size}\\
        Radial locations of substructures & Inferred from \cite{2025_Vioque} \\
        \hline
    \end{tabular}
\end{table}

\begin{table*}
\centering
\caption{Disk parameters and included substructures. Gas disk masses adopted from \citep{Trapman_2025_gas}, stellar parameters from \citep{zhang2025}. }
\label{tab:substructures}
\begin{tabular}{|l|llllll}

\hline
Disk     & \multicolumn{1}{l|}{$T_{\rm{eff}} (K)$} & \multicolumn{1}{l|}{$L_\star(L_\odot)$} & \multicolumn{1}{l|}{$M_\star(M_\odot)$} & \multicolumn{1}{l|}{$M_{g} (\log M_\odot)$} & \multicolumn{1}{l|}{$R_c$} & \multicolumn{1}{l|}{Ring/Gap location(AU)} \\ \hline
Oph1     & 3970                               & 0.9                                & 0.6                                 & -1.00                                    & 80*                     &                                      \\ \hline 
Oph2     & 3560                               & 0.4                                & 0.4                                 & -3.15                                    & 80*                     & 12 (G), 15(R), 58(R)                       \\ \hline
Oph3     & 3970                               & 0.9                                & 0.6                                 & -2.33                                    & 30*                     &                                            \\ \hline
Oph4     & 3995                               & 0.9                                & 0.6                                 & -3.40                                    & 30*                     &                                            \\ \hline
Oph5     & 3880                               & 0.8                                & 0.5                                 & -3.45                                    & 30*                     &                                            \\ \hline
Oph6     & 3425                               & 0.3                                & 0.3                                 & -1.00                                    & 30*                     & 12.5 (G), 15.5 (R)                         \\ \hline
Oph7     & 4020                               & 1                                  & 0.7                                 & -1.00                                    & 30*                     & 22 (R)                                     \\ \hline
Oph8     & 3700                               & 0.5                                & 0.4                                 & -2.64                                    & 30*                     &    25 (R)                                        \\ \hline
Oph9     & 3425                               & 0.3                                & 0.3                                 & -1.63                                    & 30*                     & 8 (G), 19 (R)                              \\ \hline
Oph10    & 3630                               & 0.5                                & 0.4                                 & -4.19                                    & 30*                     & 31.75 (R)                                  \\ \hline
Lup1     & 4060                               & 0.87                               & 0.61                                & -2.79                                    & 40                      & 27 (R)                                     \\ \hline
Lup2     & 3632                               & 0.33                               & 0.414                               & -2.07                                    & 53                      & 29 (R)                                     \\ \hline
Lup3     & 3705                               & 0.39                               & 0.47                                & -2.81                                    & 27                      &                                            \\ \hline
Lup4     & 3560                               & 0.27                               & 0.374                               & -3.87                                    & 18                      &                                            \\ \hline
Lup5     & 4060                               & 0.6                                & 0.67                                & -4.48                                    & 23                      &                                            \\ \hline
Lup6     & 3415                               & 0.2                                & 0.291                               & -3.29                                    & 25                      &                                            \\ \hline
Lup7     & 3415                               & 0.15                               & 0.301                               & -3.52                                    & 22                      &                                            \\ \hline
Lup8     & 3415                               & 0.22                               & 0.291                               & -3.59                                    & 17                      &                                            \\ \hline
Lup9     & 3415                               & 0.27                               & 0.291                               & -4.56                                    & 37                      &                                            \\ \hline
Lup10    & 4205                               & 1.21                               & 0.64                                & -1.39                                    & 273                     & **                                         \\ \hline
UppSco1  & 3700                               & 0.21                               & 0.55                                & -2.27                                    & 32                      & 47 (G), 75 (R)                             \\ \hline
UppSco2  & 3020                               & 0.07                               & 0.2                                 & -3.77                                    & 14                      &                                            \\ \hline
UppSco3  & 3490                               & 0.15                               & 0.44                                & -2.38                                    & 19                      &                                            \\ \hline
UppSco4  & 3735                               & 0.35                               & 0.672                               & -4.30                                    & 49                      &                                            \\ \hline
UppSco5  & 3360                               & 0.11                               & *                                   & -5.25                                    & 28                      &                                            \\ \hline
UppSco6  & 3700                               & 0.18                               & 0.56                                & -4.33                                    & 90                      & 12(G), 30 (R)                              \\ \hline
UppSco7  & 3490                               & 0.23                               & 0.65                                & -3.28                                    & 56                      & 12 (G), 27 (R)                             \\ \hline
UppSco8  & 3360                               & 0.14                               & 0.56                                & -2.47                                    & 40                      & 7 (G),  16 (R)                             \\ \hline
UppSco9  & 3880                               & 0.24                               & 0.61                                & -1.39                                    & 18                      & 37 (R)                                     \\ \hline
UppSco10 & 3630                               & 0.35                               & 0.6                                 & -2.74                                    & 16                      & 33 (R)                                     \\ \hline
\end{tabular}
    \begin{tablenotes}
    \small
    \item \textbf{Table notes.} The first column lists the disk name as used throughout this work. Column 2 gives the stellar effective temperature $T_{\rm{eff}}$ in Kelvin, and column 3 shows the stellar luminosity $L_\star$ in solar luminosities ($L_\odot$). Column 4 describes the stellar mass $M_\star$ in solar masses ($M_\odot$). Column 5 shows the logarithm of the disk gas mass $M_g$ in solar masses. Column 6 gives the characteristic radius $R_c$ (in AU), which is used to define the initial gas surface density profile in the model. Asterisks indicate that a fiducial value was adopted in the absence of a resolved measurement. Column 7 lists the locations of rings (R) and gaps (G) in AU, inferred from the \texttt{frank} visibility fit. **Lup 10 disks have substructures: 22 (R), 42 (G), 45 (R), 58(R), 90 (G), 95 (R), 120 (R). Disks without resolved/clear substructures are left blank and considered as smooth disks at the resolution of the AGE-PRO survey.
    \end{tablenotes}
\end{table*}

\subsubsection{Fragmentation Velocity}

The fragmentation velocity $v_{\rm{frag}}$ sets the threshold relative velocity above which colliding dust aggregates fragment rather than grow. This parameter plays a crucial role in determining the maximum grain size that particles can reach before destructive collisions become dominant \citep{2008BlumWurm}.

Laboratory experiments suggest that $v_{\rm{frag}}$ depends strongly on material properties and temperature. Silicate particles typically have low fragmentation thresholds, around $\sim 1\,\mathrm{m\,s^{-1}}$, whereas water icy grains may survive collisions up to $50\,\mathrm{m\,s^{-1}}$ under specific conditions \citep{Wada2009, Gundlach2015}. However, recent low-temperature experiments have shown that amorphous water ice may be more fragile than previously thought \citep{MusiolikWurm2019}, suggesting that even icy aggregates could have fragmentation thresholds as low as $\sim 1\,\mathrm{m\,s^{-1}}$. Motivated by these laboratory results, we model each disk using two representative values for $v_{\mathrm{frag}}$: $\sim 1\,\mathrm{m\,s^{-1}}$ and $\sim 10\,\mathrm{m\,s^{-1}}$, to bracket this range and investigate how  $v_{\rm{frag}}$ influences the resulting dust distributions. These effects are particularly relevant for understanding the supply of solids to the inner disk because radial drift depends on how big particles can grow.

\subsubsection{Turbulence Parameter}

The turbulence parameter $\alpha$ is one of the most important and uncertain parameters in the models of protoplanetary disks. It describes the efficiency of angular momentum transport and turbulent mixing in the disk, and directly influences dust diffusion, vertical settling, and collisional velocities between grains. Because these processes govern whether dust grains grow or fragment, $\alpha$ plays a central role in shaping the grain size distribution and dust evolution. In our model setup, we assume a single turbulence parameter $\alpha$  that governs both angular momentum transport and vertical dust diffusion.

Originally introduced by \citet{shakura1973black} in accretion disk theory, $\alpha$ is a dimensionless scaling factor for the turbulent viscosity $\nu_t$, such that:
\begin{equation}
\nu_t = \alpha \, c_s \, H,
\end{equation}

\noindent where $c_s$ is the local sound speed and $H$ is the gas scale height. This $\alpha$ parameter directly influences the dust radial diffusion and vertical mixing of particles in the dust evolution models.

Despite the importance of $\alpha$ for gas and dust evolution, observational constraints on $\alpha$ remain uncertain. Some studies based on dust settling suggest low levels of turbulence. For example, \cite{Villenave_2022} modeled the vertical extent of mm-sized dust grains in the highly inclined disk Oph 163131 and inferred a turbulent viscosity coefficient of $\alpha \lesssim 10^{-5}$. A recent uniform analysis of 33 disks using radiative transfer fitting of dust scale heights further supports low turbulence, finding typical upper limits of $\alpha \lesssim 10^{-3}$ across most systems \citep{Villenave2025}. However, it is noted that the $\alpha$ viscosity value inferred from vertical settling may differ from the effective $\alpha$ that governs radial transport, which further motivates the choice to explore a slightly broader range. In addition to these directly observed constraints, these values are comparable to the viscosity parameter $\alpha \sim 2-4 \times 10^{-4}$ inferred from gas evolution population synthesis modeling of the AGE-PRO sample \citep{Tabone_2025}, which reproduce the observed distributions of gas disk sizes and gas masses.

Even though analysis of dust settling from highly-inclined disks suggests that $\alpha$ could be as low as $\alpha \lesssim 10^{-5}$, we do not explore such values in this study. This is because $\alpha$ in dust evolution models does not only affect how dust settles but also sets the radial diffusion and the turbulence that limit how much grains can grow. If we set $\alpha$ too low, the grains in our dust evolution models may grow into very large particles ($>10-100 \rm{cm}$) in the pressure traps, and become invisible at sub-millimeter wavelengths due to the low opacity of these grains. This can already be seen in the case of some of our $\alpha= 10^{-4}$ simulations (see e.g. Fig \ref{fig:RADMC-3D_comparison} bottom right panel for Upp Sco 1). For this reason, we do not include $\alpha = 10^{-5}$ in the parameter space, as it is likely inconsistent with most of the structures seen in the observations. 
Therefore, we restrict our analysis to two representative values: $\alpha = 10^{-3}$ and $10^{-4}$, to investigate how the strength of turbulence affects dust growth, transport, and trapping. These values encompass typical estimates and allow us to test their effect on observable disk properties.

\subsection{Radiative Transfer with RADMC-3D}
To generate synthetic continuum observations of our dust disk models, we use the radiative transfer code \texttt{RADMC-3D} \citep{dullemond2012radmc}. First, we converted \texttt{DustPy} outputs into radiative transfer inputs for \texttt{RADMC-3D}. The 1D radial dust distribution from \texttt{DustPy} was interpolated onto a 3D spherical grid to obtain the full dust density distribution required for radiative transfer modeling. We use the vertical scale height and midplane density from \texttt{DustPy} to calculate the 3D dust density, assuming hydrostatic equilibrium.

To translate the 1D dust distribution from the dust evolution models into a 3D distribution for \texttt{RADMC-3D}, we define the spherical grid dimensions, including radial distance, co-latitude, and azimuthal angle. Next, we calculate the corresponding cylindrical coordinates ($R$, $z$) for each spherical grid cell. Using these coordinates, we interpolate the midplane density and scale height obtained from the 1D \texttt{DustPy} results onto the cylindrical radius of each grid cell. Finally, the interpolated values are expanded to fit the full 3D grid, allowing us to compute the vertical dust distribution by assuming a Gaussian profile in the vertical direction.

The dust density distribution in cylindrical coordinates is obtained by:
\begin{equation}
\rho_d(R, \phi, z) = \frac{\Sigma_d(R)}{\sqrt{2\pi} H_d(R)} \exp\left(-\frac{z^2}{2H_d^2(R)}\right),
\end{equation}
\noindent where \(R\) and \(z\) are cylindrical coordinates, and \(H_d(R)\) is the dust scale height \citep{Pohl_2016}.  $\Sigma_d(R)$ is the dust surface density, which we obtained from the \texttt{DustPy} simulation.

In our models, the dust scale height $H_{\mathrm{d}}(a)$ is computed separately for each grain size bin, assuming vertical equilibrium between sedimentation and turbulent diffusion. This follows the standard prescription:

\begin{equation}
H_{\mathrm{d}}(a) = H_{\mathrm{g}} \sqrt{\frac{\alpha}{\alpha + \mathrm{St}(a)}},
\end{equation}

\noindent where $\alpha$ is the turbulent viscosity parameter, and $\mathrm{St}(a)$ is the Stokes number of particles of size $a$. This formulation is commonly used in dust evolution models \citep{2005Dominik&Dullemond}, and is implemented in \texttt{DustPy} as the default treatment of vertical dust settling.
We note that this approach does not incorporate more complex vertical transport mechanisms, such as those described in \citep{2023Lesur}, which account for MHD-driven turbulence and vertical mixing in a more physically detailed manner. The simplified turbulent mixing prescription used here provides a reasonable approximation for the purposes of this study.

We adopt dust opacities from \cite{ricci2010dust}. This is motivated by the fact that Ricci opacities have been shown to reproduce the millimeter fluxes and spectral indices of full disk populations better than other prescriptions like the DSHARP opacities \citep[][]{Stadler2022, Delussu_2024}. Furthermore, a recent multiwavelength study of CI Tau by \citet{Zagaria_2025} compared different dust opacity prescriptions and found that the Ricci opacities provide the best agreement with observations.

The dust opacities described by \cite{ricci2010dust} use a composition of 21\% graphite, 36\% amorphous carbon and 28\% ice-water, 15\% silicate and a porosity of 40\%. The opacity files are generated using \texttt{optool} \citep{2021ascl.soft04010D}. The opacities are calculated assuming Mie theory for spherical, compact dust grains. This approach considers the complex refractive indices and provides accurate absorption and scattering properties for different dust compositions \citep{Pohl_2016}.

\subsubsection{Synthetic images}
To compare our models with ALMA observations, we compute synthetic continuum images using the \texttt{RADMC-3D}. For each dust evolution model, we generate images at ALMA Bands 3 (3.1\,mm), 5 (1.6\,mm), 6 (1.3\,mm), and 7 (0.87\,mm), using the dust density and grain size distributions from \texttt{DustPy}, and the dust opacities described before. 
We generate both face-on and inclined synthetic images. Face-on images are used to extract radial intensity profiles, consistent with the assumptions of the FRANK fits, which model disks as azimuthally symmetric and reconstruct radial profiles from deprojected visibilities. Inclined images are used for qualitative comparison with observed ALMA images, as they preserve the observed geometry and allow direct visual assessment of disk morphology.

For consistency, we convolve the inclined synthetic images with a Gaussian point spread function (PSF) matching the tclean beam, and the face-on synthetic images with a circular Gaussian PSF corresponding to the effective resolution of the FRANK reconstructions. This is tailored to each disk, which ranges from 10 to 50 AU ($\sim 0.06-0.38''$) depending on signal-to-noise and disk brightness \citep{2025_Vioque}.

Synthetic emission is analyzed using three key observational diagnostics: the radial brightness profile, which captures the spatial distribution of emission; the effective dust disk radius $R_{90\%}$, which measures the extent of the disk; and the spectral index $\alpha_{\rm{mm}}$, which provides insight into grain size and optical depth. These diagnostics are used to compare models with both FRANK-derived profiles and multiwavelength ALMA observations, allowing us to assess the ability of each dust evolution configuration to reproduce the observed disk properties.

\subsubsection{Model radial profile compared to observations}

To evaluate how well the models reproduce the observed radial structure, we extracted radial intensity profiles from the face-on or deprojected synthetic images. 

The radial distance from the center of the image is computed as:

\begin{equation}
r = \sqrt{(x - x_0)^2 + (y - y_0)^2},
\end{equation}

\noindent where $(x_0, y_0)$ is the center of the image. The pixels are then grouped by radius values of integers $r$, and their intensity values are averaged within each radial bin. This yields the radial profile:

\begin{equation}
I(r) = \frac{1}{N_r} \sum_{i \in r} I_i,
\end{equation}

\noindent where $I_i$ is the intensity at pixel $i$, and $N_r$ is the number of pixels in the annulus at radius $r$. The resulting profile is normalized by its peak for comparison with the normalized FRANK-derived profile, as we are primarily interested in how well the radial profile and substructure features (i.e. rings and gaps) are reproduced, rather than the absolute intensity scale.

We convolve the synthetic image with a Gaussian kernel matching the FRANK  uncertainty prior to extracting the profile. This ensures that resolution effects are consistent between model and data. To quantify the goodness-of-fit between the normalized model and observed profiles, we computed the reduced $\chi^2$:

\begin{equation}
\chi^2_\nu = \frac{1}{N - p} \sum_{r} \frac{\left[I_{\mathrm{RADMC}}(x, y) * G_{\mathrm{FRANK}}(x, y) - I_{\text{FRANK}}(r)\right]^2}{\sigma^2(r)},
\end{equation}

\noindent where $N$ is the number of radial bins, $p$ is the number of free parameters (in this case, zero), and $\sigma(r)$ is the estimated uncertainty in the FRANK profile at radius $r$. \( I_{\mathrm{RADMC}}(x, y) \) is the synthetic face-on intensity map from the model, and \( G_{\mathrm{FRANK}}(x, y) \) is a two-dimensional Gaussian kernel matching the FRANK resolution.

This comparison allows us to assess which model configuration best reproduces the observed profile and substructures. In addition to this quantitative comparison, we also visually inspect the synthetic and observed ALMA images to ensure that the selected models reproduce the overall emission morphology and radial extent.

\subsubsection{Dust Disk Mass Estimation}

To calculate the dust disk mass in the models, we use two approaches: directly taking the mass from the \texttt{DustPy} simulations, and deriving the mass from flux of the synthetic ALMA images assuming optically thin emission.

(i) DustPy masses:
\newline
The \texttt{DustPy} simulations trace the surface density and size distribution of dust grains over time, yielding total dust mass for each modeled disk. The masses include all grain sizes and reflect the physical dust content after dust evolution.

(ii) Dust masses from flux:
\newline
We also estimate dust masses from the synthetic continuum images at ALMA Bands 3, 5, 6, and 7 under the optically thin assumption. The dust mass is calculated using the conventional expression:

\begin{equation}
M_{\mathrm{dust}} = \frac{F_{\nu} d^2}{\kappa_{\nu} B_{\nu}(T_{\mathrm{dust}})}
\end{equation}

\noindent where $F_{\nu}$ is the flux density at frequency $\nu$, $d$ is the distance to the source, $\kappa_{\nu}$ is the dust opacity at frequency $\nu$, and $B_{\nu}(T_{\mathrm{dust}})$ is the Planck function at dust temperature $T_{\mathrm{dust}}$. 

This method assumes that dust emission is optically thin, isothermal, and dominated by large grains. We adopt a characteristic dust temperature of 20~K. We note that the optically thin assumption may underestimate the dust mass in optically thick regions. We assume the commonly used opacity values from the literature for compact grains with an MRN-like size distribution, assuming compact grains with standard astronomical silicate composition. Specifically, we use $\kappa_\nu$ = 0.9, 1.5, 2.3, and 3.4 cm$^2$ g$^{-1}$ for ALMA Bands 3, 5, 6, and 7, respectively \citep{ricci2010dust}.

\subsubsection{Effective Dust Disk Radius ($R_{90\%}$)}

The radius $R_{90\%}$ is defined as the radius at which 90\% of the total continuum flux is enclosed and is given by:

\begin{equation}
  \frac{2\pi}{F_{\text{total}}} \int_0^{R_{90}} I(r) \, r \, dr =  0.9, 
\end{equation}

\noindent where $F_{\text{total}}$ is total integrated flux and  $I(r)$ is the radial brightness profile. We calculate $R_{90\%}$ for each disk at each wavelength.

As our models are noiseless, and effectively have infinite sensitivity, we impose a 3\(\sigma\) threshold in the model image by matching the observed 3\(\sigma\) fraction of the observed peak:
\[
f_{3\sigma}\;=\;\frac{3\,\sigma_{\rm obs}}{I_{\rm pk}^{\rm obs}}, 
\qquad 
T\;=\;f_{3\sigma}\,I_{\rm pk}^{\rm mod},
\]
\noindent where T is the intensity threshold, and \(\sigma_{\rm obs}\) is the observed RMS (mJy/beam), \(I_{\rm pk}^{\rm obs}\) the observed peak
(mJy/beam), and \(I_{\rm pk}^{\rm mod}\) the peak of the convolved model image (mJy/beam).
We retain pixels if they satisfy \(I_{\rm mod}(x,y)\ge T\)
and \(R_{90}\) is then measured from the masked image.

\begin{figure*}
    \centering
    \includegraphics[width=0.96\textwidth]{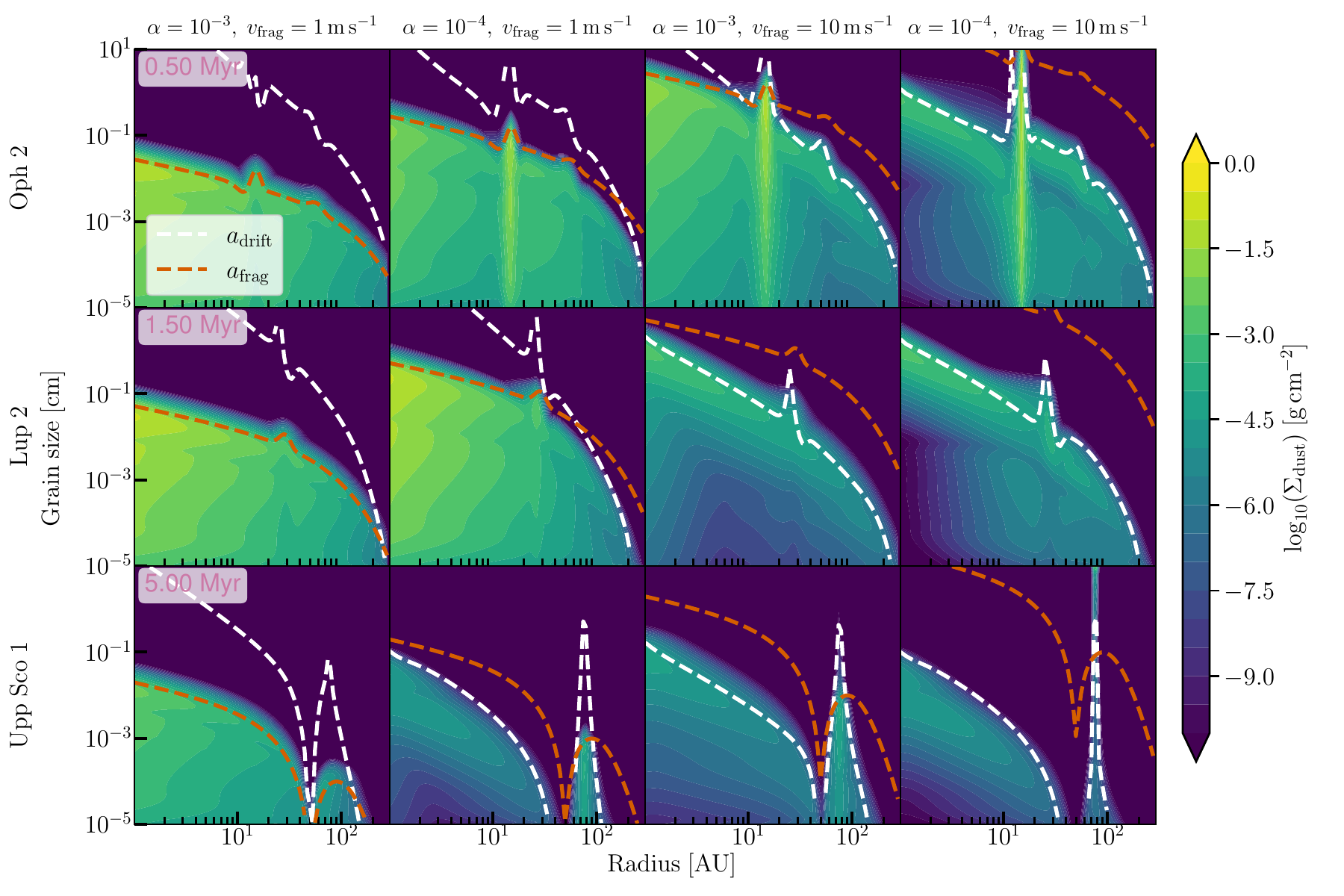}
    \caption{Dust surface density distributions from \texttt{DustPy} simulations for three representative disks — Oph 2 (top row), Lup 2 (middle row), and Upp Sco 1 (bottom row). Each column corresponds to a different model configuration, varying the turbulence parameter $\alpha$ and fragmentation velocity $v_{\mathrm{frag}}$. The red dashed line shows the fragmentation limit ($a_{\mathrm{frag}}$) and the white shows the drift limit ($a_{\mathrm{drift}}$) at each radius. Models with low $\alpha$ and high $v_{\mathrm{frag}}$ (rightmost column) allow the most efficient grain growth and trapping, while high $\alpha$ and low $v_{\mathrm{frag}}$ (leftmost column) lead to smaller grains and inefficient dust trapping.}
    \label{fig:dustevolution}
\end{figure*}

\subsubsection{Spectral Index ($\alpha_{\rm{mm}}$)}
The spectral index between two observing bands provides insight into grain size distributions and optical depth, and between two wavelengths is given by:
\begin{equation}
    \alpha_{\rm{mm}} = \frac{\log(F_{\nu_1} / F_{\nu_2})}{\log(\nu_1 / \nu_2)} = -\frac{\log(F_{\nu_1} / F_{\nu_2})}{\log(\lambda_1 / \lambda_2)},
\end{equation}

\noindent where $F_{\nu_1}$ and $F_{\nu_2}$ are the flux densities at frequencies $\nu_1$ and $\nu_2$ and $\lambda_1$ and $\lambda_2$ are the wavelengths corresponding to $\nu_1$ and $\nu_2$.

A value of $\alpha_{\rm{mm}}\sim2-3$ typically indicates large grains, while lower values indicate optically thick emission or contributions from non-thermal emission mechanisms \citep[e.g., free-free or synchrotron][]{Testi_2014, 2014A&A...564A..51P, rota2024}. Higher values of $\alpha$ suggest that the grains are small as the interstellar medium, i.e., predominantly micrometer-sized particles. Variations in $\alpha_{\rm{mm}}$ with radius can thus reflect changes of the radial grain size distribution and/or opacity gradients. 

By comparing synthetic images at Bands 3, 5, and 6, we assess how the spatial distribution of emission changes with wavelength, which traces the underlying grain size distribution. Longer wavelengths are more sensitive to larger grains and can reveal whether growth and drift processes preferentially concentrate particles in specific disk regions or structures.
These synthetic diagnostics are compared to FRANK-derived profiles, observed ALMA images, and AGE-PRO results in the following section, allowing us to assess the physical viability of each model configuration.

\subsection{Estimation of water mass to compare with JWST observations}

To assess the potential delivery of cold water into the inner disk, we estimate the mass of cold water ice using the method introduced by \citet{romeromirza2024retrievalthermallyresolvedwatervapor}. The mass of cold H$_2$O delivered via pebble drift is given by:

\begin{equation}
    M[\text{Cold H}_2\text{O}] \approx \dot{M}_{\text{peb}} \, t_{\text{H}_2\text{O}} \, \frac{f_{\text{ice}} f_{\text{H}_2\text{O}}}{\eta},
    \label{eq:watermass}
\end{equation}

\noindent where $\dot{M}_{\text{peb}}$ is the pebble flux. Following \citet{romeromirza2024retrievalthermallyresolvedwatervapor}, we adopt fiducial values of $f_{\text{ice}} = 0.2$, representing the fraction of pebble mass composed of volatile ices, and $f_{\text{H}_2\text{O}} = 0.8$, assuming water ice is the dominant component of the ice fraction. The efficiency factor $\eta = 10^3$ accounts for the uncertainty in translating the pebble delivery rate into a detectable water vapor mass, encompassing losses due to sublimation, photodissociation, accretion onto planets, or inefficient conversion into gas-phase H$_2$O. The timescale $t_{\text{H}_2\text{O}} = 100$ yr approximates the residence time of water vapor in the inner disk, based on chemical modeling and replenishment rates constrained by JWST observations. We note that the factors involved in Eq.~\ref{eq:watermass} are highly uncertain \citep[e.g.,][]{houge2025}, and we adopt the same quantities as \citet{romeromirza2024retrievalthermallyresolvedwatervapor}.

The pebble flux $\dot{M}_{\text{peb}}$ is evaluated at the snowline or 1 AU, serving as a proxy for the inward delivery rate of solids into the warm inner disk. As such, this gives us a first-order estimate of the mass of water ice delivered via drifting pebbles.

\section{Results}  \label{sect:results}

We present the results of dust evolution modeling and radiative transfer for the 30 disks in the AGE-PRO sample, spanning the star-forming regions Ophiuchus, Lupus, and Upper Scorpius. 

To illustrate the model results, we will first highlight one representative disk from each star-forming region (Figures \ref{fig:dustevolution}--\ref{fig:spectralR90}). We then present full-sample statistical results across all 30 disks in Sections~4.2--4.5 (Figures ~\ref{fig:Violin_per_bands}--\ref{fig:Coldwater_substructure_rad}).

Since individual stellar ages are uncertain, we adopt representative times to extract the model outputs: 0.5 Myr for Ophiuchus, 1.5 Myr for Lupus, and 5 Myr for Upper Scorpius. These snapshots approximately reflect the evolutionary stage of each region and are used to extract model output at corresponding times for consistent comparison. 

To illustrate the model results, we will first highlight one representative disk from each region. For each selected disk, we present:
\begin{itemize}
    \item The dust surface density distribution and maximum grain size from DustPy
    \item Synthetic ALMA continuum images from RADMC-3D 
    \item Multiwavelength radial profiles and spectral index profile, in addition to the $R_{90\%}$ for each wavelength.
\end{itemize}

These individual cases serve to highlight the range of dust evolution outcomes across different turbulence and fragmentation regimes and their fit to observations, and for the rest of the disks we only give the best combination of $\alpha$ and $v_{\rm{frag}}$. 

We then transition to the full AGE-PRO sample to quantify population-level trends, comparing the modeled disk masses, effective radii ($R_{90\%}$), and spectral indices to the results of AGE-PRO. Finally, we examine how well each model matches the observed radial profiles, and quantify the resulting inner disk pebble fluxes as a potential indicator of the amount of inner disk water vapor, with the prospect of comparing to JWST observations.

\subsection{Representative disks: Oph 2, Lup 2, and Upp Sco 1}

We select one representative disk from each star-forming region: Oph 2 (Ophiuchus), Lup 2 (Lupus), and Upp Sco 1 (Upper Scorpius). Each disk was modeled using the four outlined configurations, with the turbulence parameter ($\alpha = 10^{-4},\ 10^{-3}$) and the fragmentation velocity ($v_{\mathrm{frag}} = 1,\,10\,\mathrm{m\,s^{-1}}$).

Figure~\ref{fig:dustevolution} presents the resulting dust grain size and surface density distributions for the representative disks. The outcomes reveal a clear dependence of grain growth and retention on both turbulence strength ($\alpha$) and fragmentation velocity ($v_{\mathrm{frag}}$).

This behavior is consistent with the theoretical expectation for the fragmentation-limited grain size, which scales as:
\begin{equation}
a_{\mathrm{frag}} \propto \frac{v_{\mathrm{frag}}^2}{\alpha},
\end{equation}

\noindent indicating that grain growth to larger sizes is favored by lower levels of turbulence and higher fragmentation thresholds. However, the maximum grain size can also be limited by drift, in which case the maximum grain size is given by:

\begin{equation}
a_{\mathrm{drift}} = \frac{2}{\pi} \cdot \frac{\Sigma_{\mathrm{d}}}{\rho_{\mathrm{s}}} \cdot \frac{v_{\mathrm{K}}^2}{c_{\mathrm{s}}^2} \cdot \left| \frac{d \ln P}{d \ln r} \right|^{-1},
\end{equation}

\noindent where $\Sigma_{\mathrm{d}}$ denotes the dust surface density, $\rho_{\mathrm{s}}$ is the volume density of the dust grains, $v_{\mathrm{K}}$ is the Keplerian velocity, $P$ is the gas pressure, and $r$ is the radial coordinate, and $c_{\mathrm{s}}$ is the isothermal sound speed of the gas.

Consistent with this scaling, the model with low turbulence and high fragmentation velocity ($\alpha = 10^{-4}$, $v_{\mathrm{frag}} = 10\,\mathrm{m\,s^{-1}}$
) yields the most efficient growth, producing mm–cm-sized grains that remain concentrated in the outer disk due to enhanced dust trapping. In contrast, the combination of high turbulence and low fragmentation velocity ($\alpha = 10^{-3}$, $v_{\mathrm{frag}} = 1\,\mathrm{m\,s^{-1}}$) results in significantly smaller maximum grain sizes. These smaller particles couple more strongly to the gas and are less efficiently trapped, leading to more diffuse dust distributions.

\begin{figure}
    \centering
    \includegraphics[width=\columnwidth]{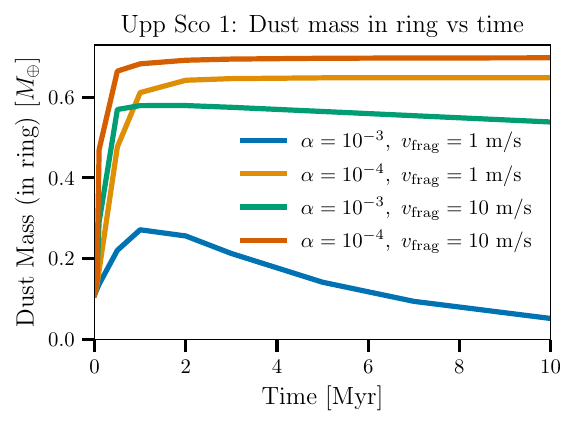}
    \caption{Evolution of the dust mass within the 75$\pm$10~AU ring of Upp Sco 1 over time, shown for the four model configurations combining turbulence parameter $\alpha = 10^{-4}, 10^{-3}$ and fragmentation velocity $v_{\rm frag} = 1, 10\,\mathrm{m\,s^{-1}}$. We illustrate how different disk evolution parameters affect local dust retention and depletion. Models with lower $\alpha$ and higher $v_{\rm frag}$ sustain higher dust mass over longer timescales, highlighting more efficient grain growth and trapping.  }
    \label{fig:Dust_mass_Upp_Sco_ring}
\end{figure}

In the case of Upp Sco 1, dust trapping occurs in all models. However, only the low-$\alpha$ and high-$v_{\mathrm{frag}}$ configurations retain a significant population of mm-sized grains in the outer ring that produce detectable continuum emission at ALMA wavelengths. This is also illustrated in Fig.~\ref{fig:Dust_mass_Upp_Sco_ring}, where the mass of the dust inside the ring is plotted against time for each value of $\alpha$ and $v_{\rm{frag}}$.

In addition to its dependence on $\alpha$ and $v_{\rm{frag}}$, Figure~\ref{fig:dustevolution} also shows a clear age dependence: older disks (e.g., Upp Sco 1 at 5 Myr) exhibit stronger depletion of mm–cm-sized grains in the outer disk, reflecting the cumulative effect of radial drift.

\begin{figure*}
    \centering
    \includegraphics[width=\textwidth]{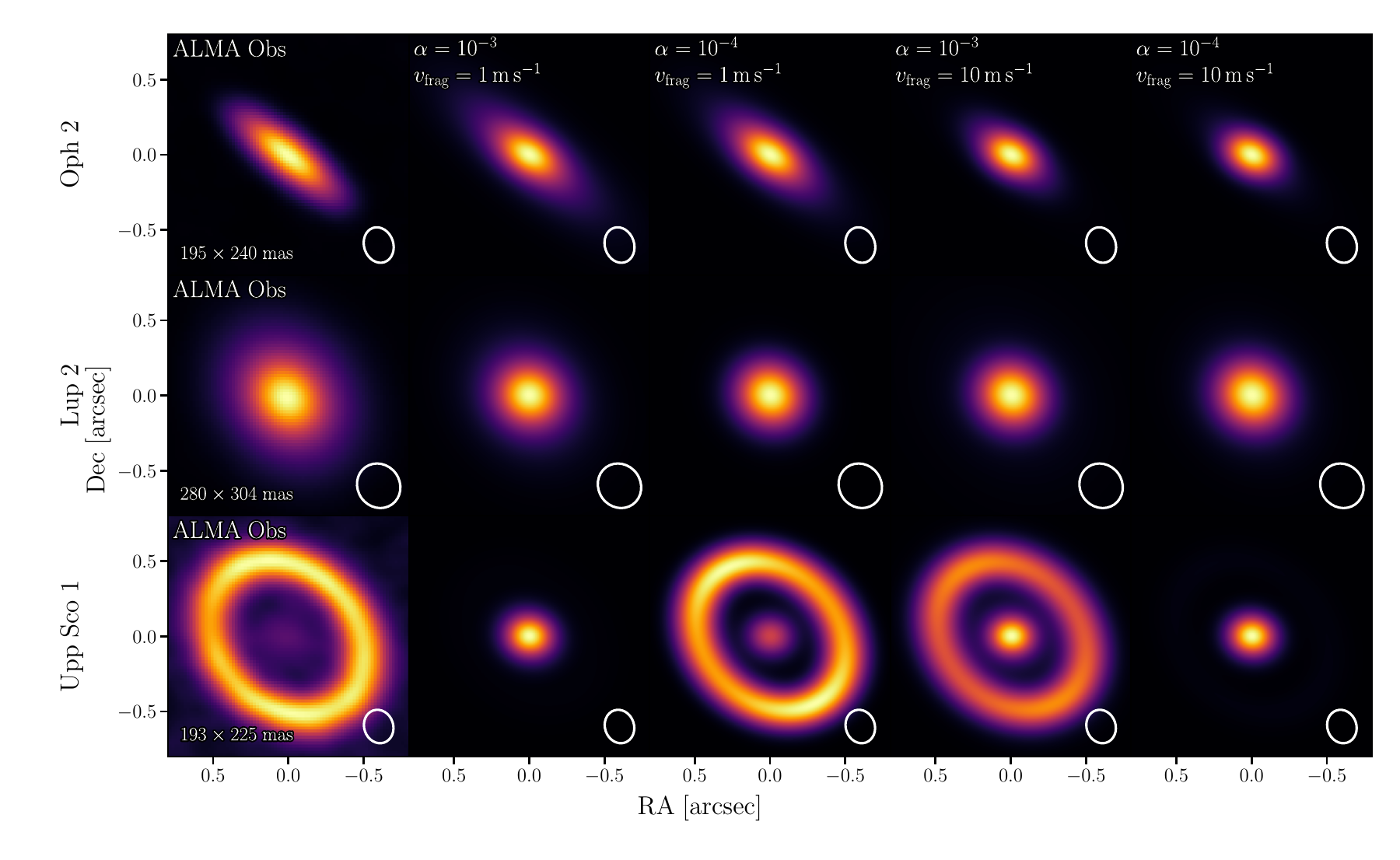}
    \caption{
    Comparison of synthetic Band 6 continuum images (RADMC-3D) with ALMA observations for Oph 2, Lup 2, and Upp Sco 1. Each row corresponds to one disk, and the columns show ALMA observations followed by synthetic images from four model configurations varying in $\alpha$ and $v_{\mathrm{frag}}$. The spatial resolution and beam size are matched with the ALMA observation. }
    \label{fig:RADMC-3D_comparison}
\end{figure*}

Figure~\ref{fig:RADMC-3D_comparison} presents synthetic Band 6 (1.3 mm) continuum images generated with RADMC-3D for the same three representative disks. The synthetic images are convolved to match the resolution of the ALMA observations. In the cases of Oph 2 and Lup 2, identifying the best-fit model from the Band 6 continuum images alone is not straightforward, as all four model configurations yield similar morphologies at the resolution of the data. In both systems, the synthetic emission generally appears more compact than observed. This discrepancy likely arises from the efficiency of radial drift in the models, which concentrates large grains in a narrow region near the pressure bump. For example, in Oph 2, the visibility fitting reveals a prominent ring at approximately 15 AU \citep{2025_Vioque}, which we include in our modeled gas profile. As particles drift and accumulate efficiently at this single trap, the outer disk becomes depleted of mm-sized grains, limiting the ability of the model to reproduce the observed extended emission (see Fig.~\ref{fig:RADMC-3D_comparison}).

However, in the case of Upp Sco 1, the best-fitting model is easier to identify, as the effects of dust trapping and grain growth leave clearer signatures in the emission morphology. The presence of a prominent ring allows us to test how dust trapping and grain growth interact. Only the models with either low turbulence and low fragmentation velocity ($\alpha = 10^{-4}$, $v_{\mathrm{frag}} = 1\,\mathrm{m\,s^{-1}}$) or high turbulence and high fragmentation velocity ($\alpha = 10^{-3}$, $v_{\mathrm{frag}} = 10 \,\mathrm{m\,s^{-1}}$) reproduce sufficient mm-sized grains in the outer disk to match the ALMA morphology. In the high-$\alpha$, low-$v_{\mathrm{frag}}$ case, grains remain small and are less efficiently trapped, while in the low-$\alpha$, high-$v_{\mathrm{frag}}$ case, growth is so efficient that much of the mass shifts to cm-sized particles, which emit less efficiently at 1.3 mm.

All model configurations produce a relatively bright inner disk component, which appears more prominent than in the observations. While an inner disk is indeed detected in the data, it is fainter than in the models. This suggests that  the trap implemented in the models is too leaky—allowing excessive inward transport of dust. Among these models, the case with $\alpha=10^{-4}$ and $v_{\mathrm{frag}} = 1\,\mathrm{m\,s^{-1}}$ leads to the best match between models and observations.

\subsubsection{Multiwavelength analysis and comparison with ALMA observations}

\begin{figure}
    \centering
    \includegraphics[width=0.45\textwidth]{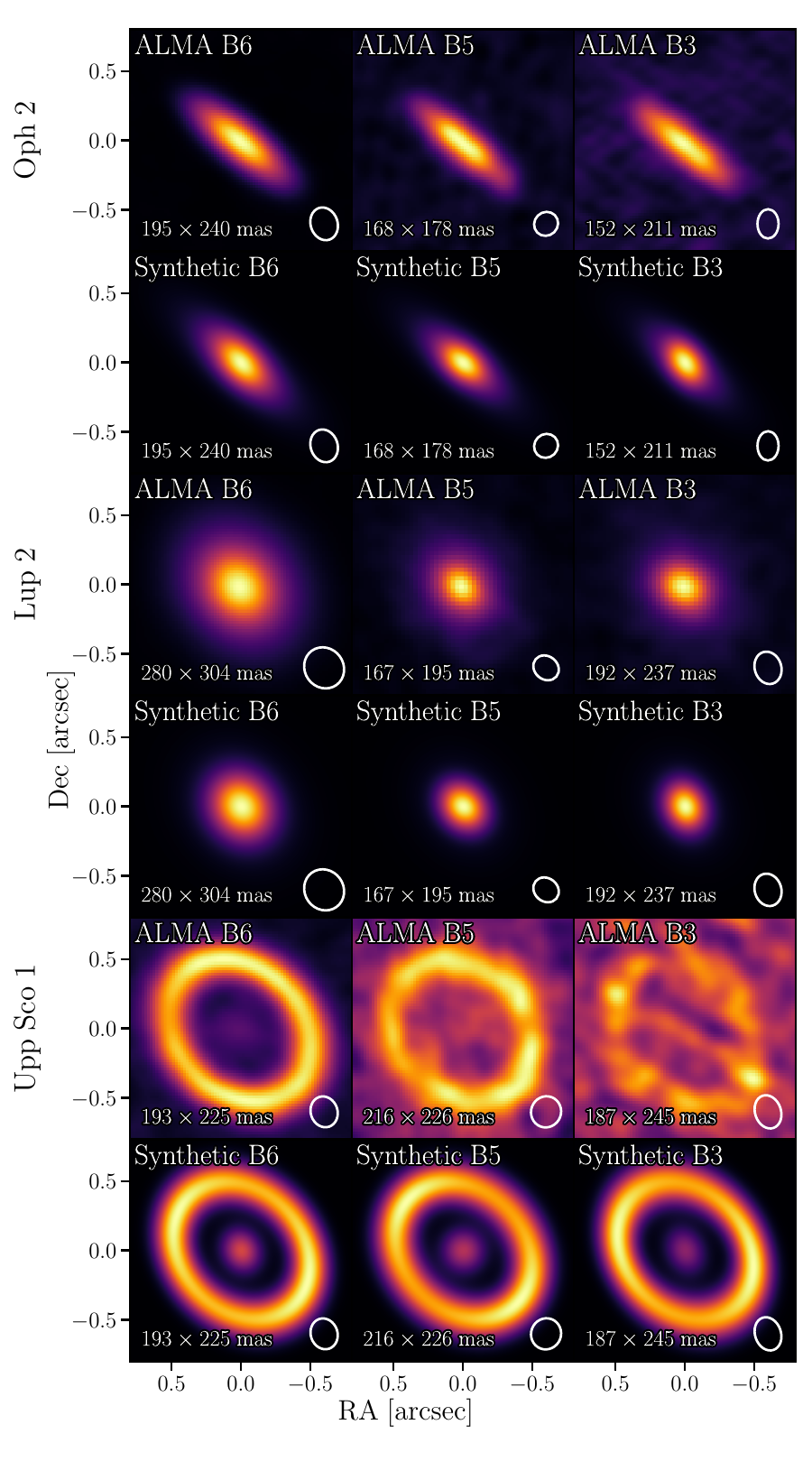}
    \caption{Multiwavelength comparison between ALMA observations and synthetic continuum images for three representative disks — Oph 2, Lup 2, and Upp Sco 1. The top panels show ALMA observations at Bands 6 (1.3 mm), 5 (1.6 mm), and 3 (3.0 mm). The bottom panels show the best-fit synthetic images for each disk, corresponding to the combination of $\alpha$ and $v_{\rm{frag}}$ that best reproduces the observed radial intensity profile at 1.3 mm. All synthetic images are convolved with the appropriate beam to match the ALMA resolution.}
    \label{fig:multiwavelength}
\end{figure}

Given the availability of multiwavelength observations for the AGE-PRO sample, we also examine and compare the synthetic continuum emission at multiple ALMA wavelengths. This comparison is key for probing the radial distribution of grain sizes, as different wavelengths are sensitive to different particle populations.

Figure~\ref{fig:multiwavelength} presents a side-by-side view of ALMA Band 6, Band 5, and Band 3 observations (top panels), and synthetic images generated from our best-fit models at the same bands (bottom panels), for the three disks. The images have been normalized to the peak flux. We find that the wavelength-dependent morphology of the synthetic images broadly reproduces the observed trend of increasingly compact emission at longer wavelengths. This behavior is expected, as longer wavelengths probe larger grains, which tend to drift more efficiently to regions of high pressure and make the disk appear less extended at longer wavelengths. This is clearer in the Oph 2 and Lupus 2 disks.

\subsubsection{$R_{90\%}$ at different wavelengths }
\begin{figure*}
    \centering
    \includegraphics[width=0.96\textwidth]{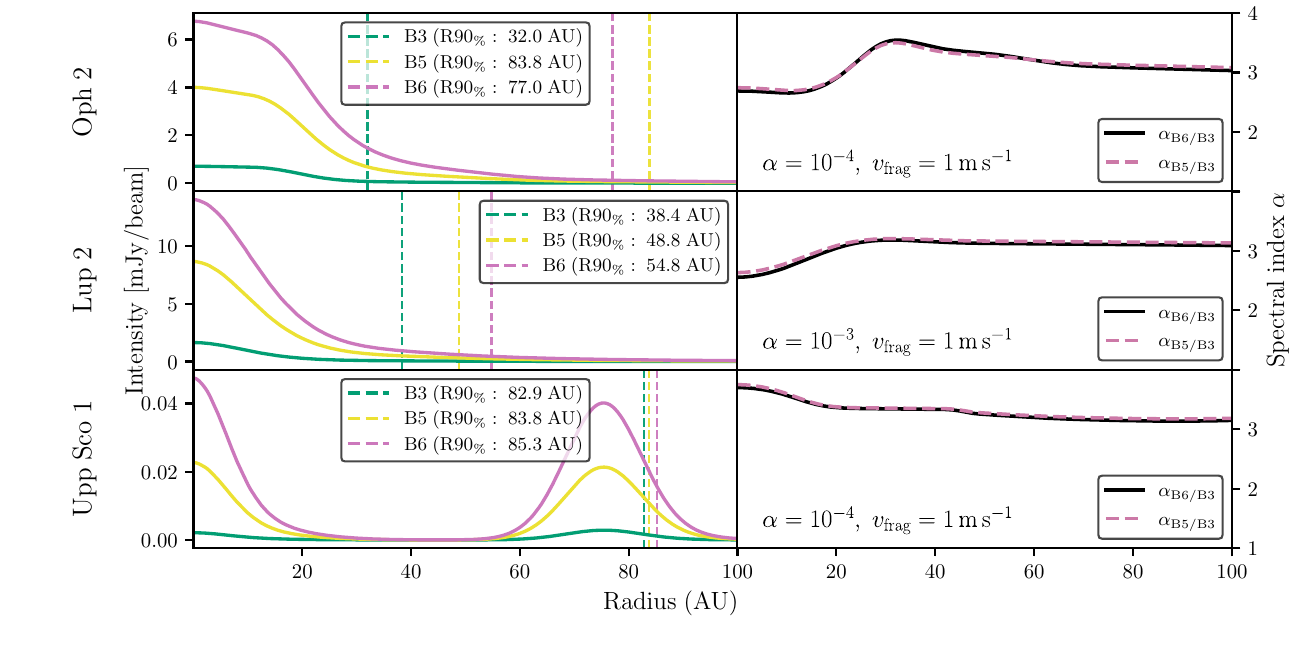}
    \caption{Radial intensity profile (left panels) and spectral index radial profile (right panels) of the selected disks from Oph, Lup and Upp Sco region for the best combination of $\alpha$ and $v_{\rm{frag}}$ for each disk. In the left panels (radial intensity profile) purple shows band 6 (1.3 mm), yellow band 5 (1.6 mm), and green band 3 (3.0 mm). Vertical dashed lines indicate the $R_{90\%}$ values for each ALMA band, marking the radius enclosing 90\% of the total continuum flux at each wavelength. In the right panels, we show the radial spectral index between Band 6 and Band 3 in black, and between Band 5 and Band 3 in purple.
}
    \label{fig:spectralR90}
\end{figure*}

To quantify how grain size distribution affects the radial extent of the continuum emission, we compute $R_{90\%}$ — the radius enclosing 90\% of the total flux — for synthetic images at Bands 3, 5, and 6. Figure~\ref{fig:spectralR90} shows the radial intensity profiles and $R_{90\%}$ values for three representative disks at each wavelength.

For Oph 2 and Lup 2, we observe a consistent decrease in $R_{90\%}$ with wavelength: from 83.8 to 32 AU in Oph 2, and from 35 to 29 AU in Lup 2. This trend aligns with expectations from radial drift, where larger grains (traced by longer wavelengths) are more centrally concentrated, while smaller grains remain distributed over larger radii.
In contrast, Upp Sco 1 shows nearly identical $R_{90\%}$ values ($\sim$83-85 AU) at all three wavelengths. This wavelength-independent $R_{90\%}$ is a direct result of the strong pressure trap imposed in the model, which retains grains of all sizes at the same location.

\subsection{Full-sample results}

\begin{figure*}
    \centering
    \includegraphics[width=0.96\textwidth]{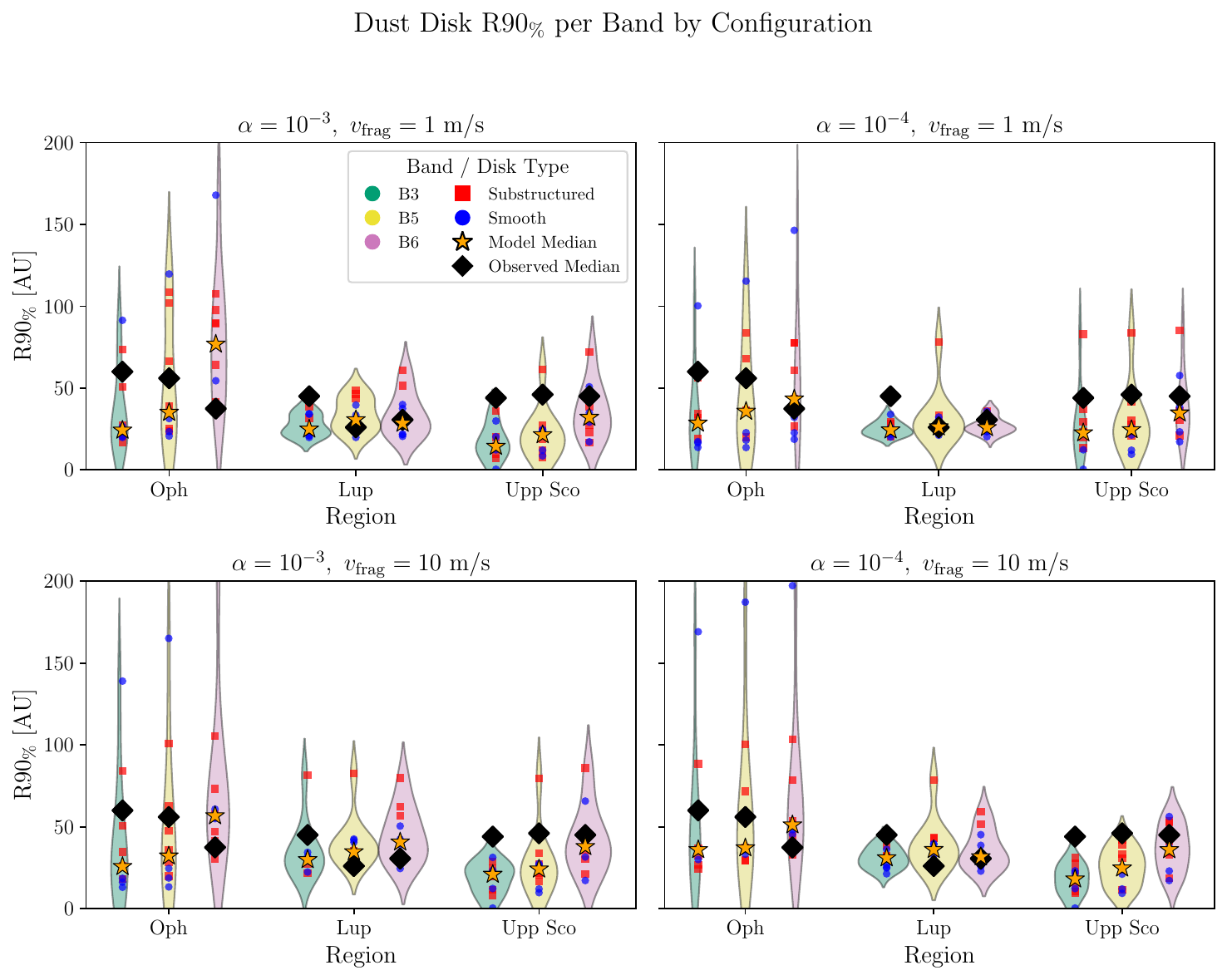}
    \caption{Distributions of $R_{90\%}$ in modeled disks per band for each combination of $\alpha$ and $v_{\rm{frag}}$, shown separately for the three star-forming regions. Points show individual modeled disks, with smooth disks in blue and disks classified as substructured in red. The median of the model $R_{90\%}$ distribution is indicated by a yellow star and the observed medians from the AGE-PRO sample are denoted with a diamond (where Band 6 median is taken from \protect\cite{zhang2025} while Band 5 and 3 medians are from Pulgarés et al, in prep). Disks labeled as "Substructured" are those with observed radial substructures (as detected in the observations by \protect\cite{2025_Vioque}) in Band 6, while "Smooth" disks show no clear substructure at the resolution of the observations.The width of each violin in the plot reflects the kernel density estimate (KDE) of the $R_{90\%}$ distribution; wider regions correspond to higher probability density.}
    \label{fig:Violin_per_bands}
\end{figure*}

Figure~\ref{fig:Violin_per_bands} shows the distribution of $R_{90\%}$ for all the modeled disks in each region and for each combination of $\alpha$ and $v_{\rm{frag}}$. Overall and independent of the value of $\alpha$ and $v_{\rm{frag}}$, we do not find significant or systematic evolution of $R_{90\%}$ with the age of the region. This is in agreement with the results from \cite{2025_Vioque}, which has been interpreted as the result of having pressure bumps in the disks and keeping the dust disk size approximately constant with time \citep{Kurtovic2025}.

Nevertheless, when examining the time evolution of $R_{90\%}$ in more detail across Oph, Lup, and Upper Sco, we do find some small trends emerge. The median continuum size  decreases with age across all $\alpha$, $v_{\rm frag}$ configurations and bands from Oph to Upper Sco, although this decrease is not always present between consecutive regions (Oph–Lup or Lup–Upper Sco). The trend is monotonic for Bands 3 and 5 in the $\alpha=10^{-4}$ models (both $v_{\rm frag}$, and for Band 5 in the $\alpha=10^{-3}$ models, while the Band 6 trends are weaker and not strictly monotonic.

For Ophiuchus and Upper Sco regions, we find that $R_{90\%}$ increases with observing frequency from Band 3 to Band 6. We do not observe this trend for the Lupus region, except for the $\alpha=10^{-3}$, $v_{\rm{frag}}=10$m\,s$^{-1}$ case. For the smooth disks and disks with weak traps, in general we find that our model $R_{90\%}$ increases with observing wavelength from Band 3 to Band 6 — consistent with expectations from dust evolution in the presence of grain growth and drift. This is also consistent with the findings of \cite{2019Rosotti} who showed that the observed continuum can be set by an opacity effect and therefore be wavelength dependent, since dust opacity decreases towards longer wavelengths. \cite{2019Rosotti} finds that disk radii measured from sub-millimeter continuum observations do not trace the true mass extent or the sharp outer edge of the dust distribution, but instead marks the radius beyond which the dust grains become too small to provide a significant sub-millimeter opacity, causing the emission to fall below detectability.

In the disks with larger traps we find that there is no significant variations of $R_{90\%}$ across wavelength, suggesting that $R_{90\%}$ is largely independent of the grain sizes. This is also in agreement with the scenario in which pressure bumps maintain a similar dust disk size, as illustrated for Upper Sco 1 in Fig.~\ref{fig:spectralR90}, and with the observed trends of $R_{90\%}$ in Pulgarés et al. (in prep). Overall, the model and observed median $R_{90\%}$ values are only partially in agreement: we find the closest correspondence for Lupus region in bands 5 and 6, with larger differences being found in other bands and regions. Given the intrinsic scatter of both the models and observations, we find that the medians are broadly consistent, within a factor of a few. However, our models do not systematically agree with the observations.

Based on the  $R_{90\%}$ values at each  wavelength and in each star-forming regions, there is not a clear favored combination of $\alpha$ and $v_{\rm{frag}}$ that can explain the observations. We find that the inclusion of substructures (highlighted in red) makes the dust disk size more extended in general.  

\subsubsection{Spectral index}

\begin{figure*}
    \centering
    \includegraphics[width=0.96\textwidth]{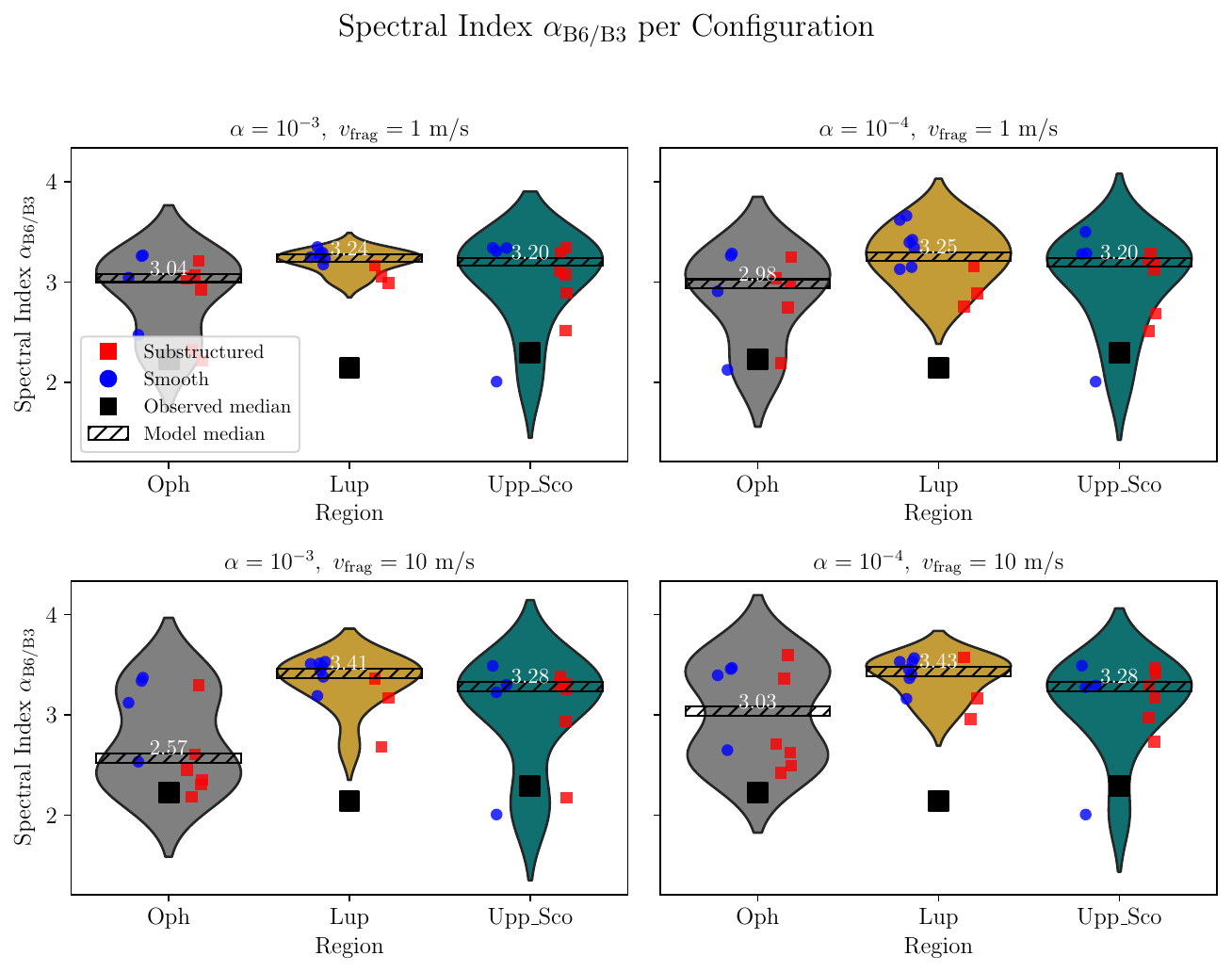}
    \caption{Spectral index B6/B3 per region for the combinations of $\alpha$ and $v_{\rm{frag}}$. The scatter shows the smooth disks in blue, the substructured disks in red and the model median is the horizontal hatched black line. Observational median values from the AGE-PRO sample are denoted as black squares (Pulgarés et al., in prep). The width of each violin in the plot reflects the kernel density estimate (KDE) of the distribution of model dust masses; wider regions correspond to higher probability density. Here we have moved the horizontal positions of substructured and smooth disks slightly along the x-axis to better help distinguish between the two groups. The horizontal positions of individual points have no physical meaning.} 
    \label{fig:spectralB}
\end{figure*}

Figure~\ref{fig:spectralB} shows the integrated spectral index calculated for each disk between 1.3\,mm (Band 6) and 3.0\,mm (Band 3) and it is shown for each combination of $\alpha$ and $v_{\rm{frag}}$. We find that in each of the models with different $\alpha$ and $v_{\rm{frag}}$, the trend is similar with lower values of $\alpha_{\rm{mm}}$ for the youngest disks in Ophiuchus compared to the the older disks in Lupus and Upper Sco. This could be the result of two non-exclusive effects: (1) particles being lost over time towards the star due to drift, increasing the spectral index,  (2) the disks becoming more optically thin with time. Overall, the spectral indices from the models are higher than observed. In the observations, some of the disks have spectral index lower than two, which is not the case for any of our models. This could be because our radiative transfer models do not include any contributions from non-thermal emission, such as free-free that can decrease the spectral index values. We note that  we do include full-scattering on our models, so it cannot be the reason for discrepancy.

\begin{figure*}
    \centering
    \includegraphics[width=0.96\textwidth]{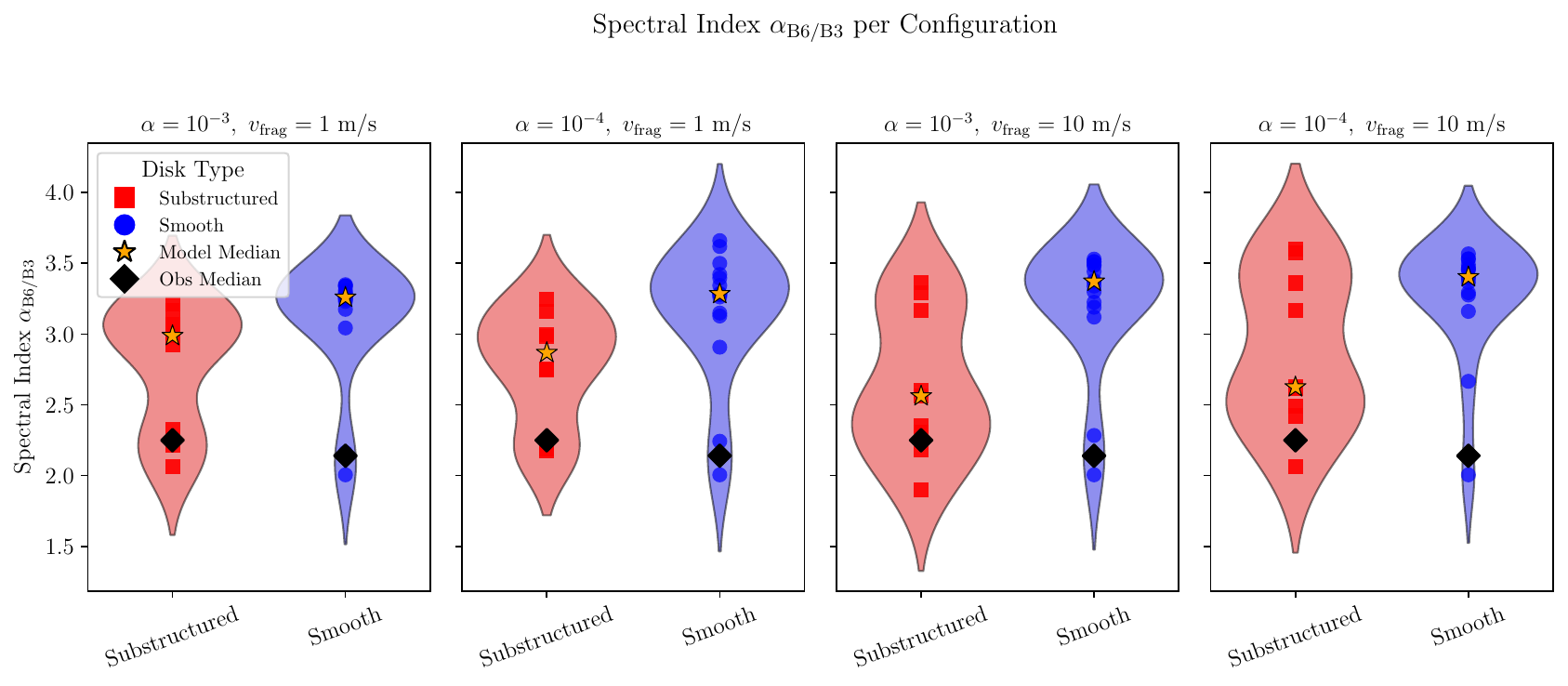}
    \caption{We present integrated spectral index $\alpha_{\rm{mm}}$ B6/B3 for smooth and substructured disks for the combinations of $\alpha$ and $v_{\rm{frag}}$. The model medians (denoted by a star) are compared with observational values (denoted as a black square) from the AGE-PRO sample. Disks labeled as “Substructured” are those where radial substructures (e.g., rings or gaps) are observationally detected in the Band 6 visibility modeling \citep{2025_Vioque}, while “Smooth” disks show no clear substructure at the resolution of the observations.}
    \label{fig:Smooth_vs_Sub_B6B3}
\end{figure*}

We also assess the variations of the spectral index for our substructured and smooth disks for each combination of $\alpha$ and $v_{\rm{frag}}$ (Fig.~\ref{fig:Smooth_vs_Sub_B6B3}) and compare them with the median observed values. Overall, the spectral index of the structured disks is slightly lower than that of the smooth disks, in agreement with previous dust evolution models that include pressure bumps \citep{pinilla2012}. Pulgarés et al. (in prep) find that the spectral index is higher for the substructured disks, opposite to the expectations from our models. The discrepancy may be an effect of optical depth. The results of Pulgarés et al. (in prep.) suggest many disks may be optically thick at (sub)millimeter, with some disks remaining partially optically thick even at band 3, leading to a lower observed spectral index and a larger emission radius. In contrast, our models are predominantly optically thin at millimeter wavelengths, leading to higher $\alpha_{mm}$ and more compact brightness profiles.

Within the substructured disk sample, we examine variations of the spectral index between gaps and rings at the resolution of FRANK. Specifically, we integrate over an area defined by the FRANK beam centered around the peak of the ring or the minimum of the gap. We do not find significant variations of $\alpha_{\rm{mm}}$ between these two locations. The test was motivated by Pulgarés et al. (in prep), who find differences in $\alpha_{\rm{mm}}$ between these two locations in the observational data. This could be a result of insufficient resolution to detect the possible variations of $\alpha_{\rm{mm}}$ that are seen for example in Fig.~\ref{fig:spectralR90}.

\subsection{Dust Disk Mass Distribution} \label{sect:results_dust_disk_mass}

\begin{figure*}
    \centering
    \includegraphics[width=0.96\textwidth]{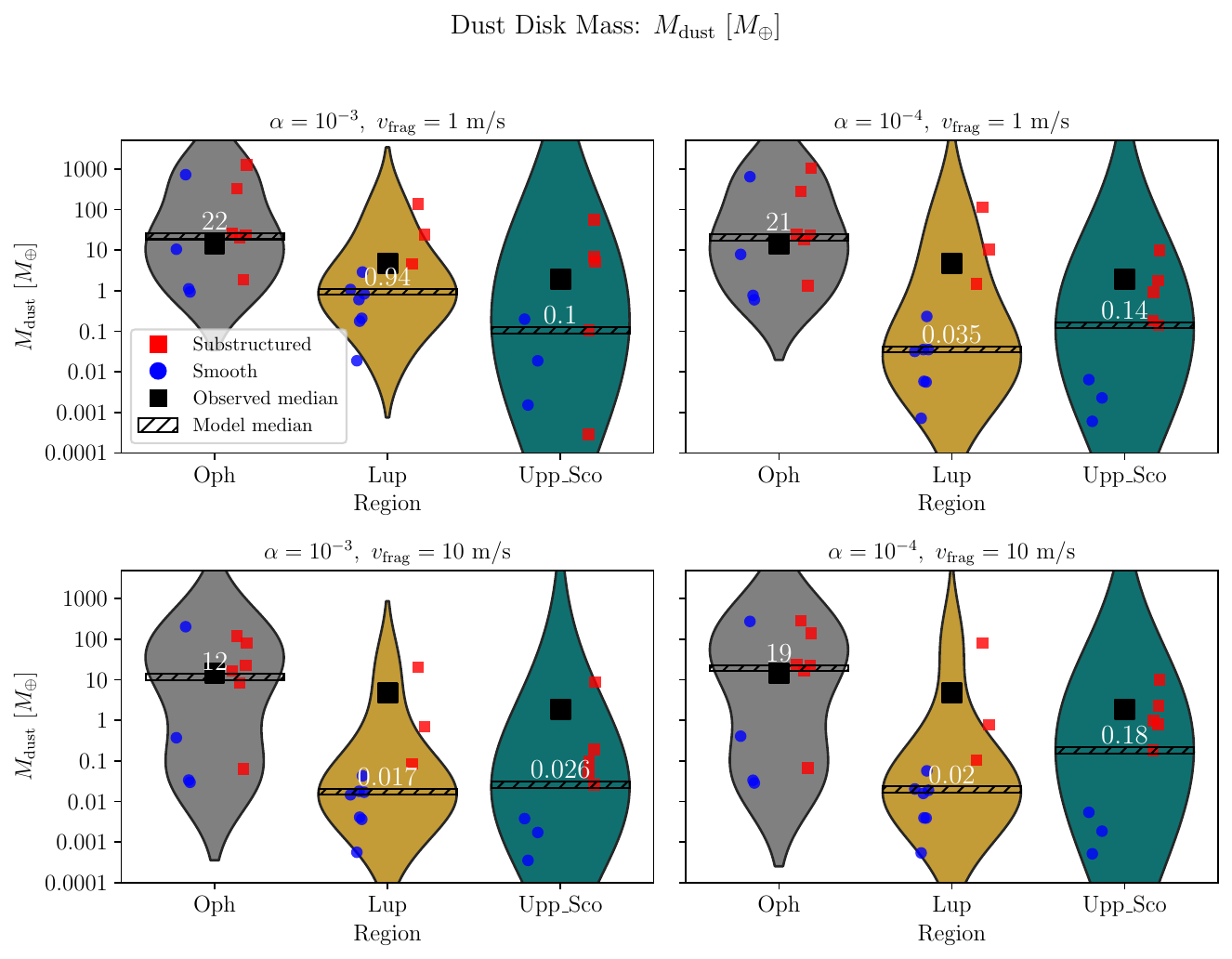}
    \caption{
        Dust mass ($M_{\mathrm{dust}}$) from Ophiuchus, Lupus, and Upper Sco regions for the combinations of $\alpha$ and $v_{\mathrm{frag}}$. Red dots represent individual substructured disks, and blue dots represent smooth disks. The black squares represent the median values obtained from the AGE-PRO observations, while the horizontal hatched line shows the median from the models. Disks are labeled as “Substructured” if radial features such as rings or gaps are detected in the visibility profiles, and as “Smooth” if no clear features are resolved at the available observational resolution.}
    \label{fig:Mdust_violin}
\end{figure*}

Fig. \ref{fig:Mdust_violin} illustrates the distribution of dust disk masses ($M_{\text{dust}}$) in logarithmic scale. The dust disk masses are directly from the models and is calculated using all grain sizes. We compare these dust disk masses with those calculated from the synthetic observations at all bands under the optically thin assumption, and find that,  independent of wavelength, the synthetic observation recover the dust disk mass from the models overall well.  In the figure, the squares represent the median values obtained from the AGE-PRO observations \cite{zhang2025}, while the horizontal hatched lines and the values above them represent the median values from the models. 

Across all models, the Ophiuchus disks are significantly more massive than Lup and Upper Sco, showing that at the age of Ophiuchus the drift has not removed as much dust particles compared with the older regions. This is consistent with what we find in the spectral indices (see fig. \ref{fig:spectralB}). The decrease in the median observational dust mass across regions is reproduced only in the models with $\alpha=10^{-3}$ and $v_{\mathrm{frag}} = 1\,\mathrm{m\,s^{-1}}$. In the other models a clear decrease can be seen from Ophiuchus to Lupus, but not from Lupus to Upper Sco.

In Fig~\ref{fig:Mdust_violin}, we compare the dust masses predicted by the model for each individual disk with the median of the observed dust masses obtained from band 6 in each corresponding star-forming region. When comparing with the observed medians we underestimate the dust masses, particularly in Lupus and Upper Sco. Underestimation of the dust mass usually occurs in the smooth disks that do not exhibit clear substructures. This, in conjunction with the low spectral indexes measured in these disks \citep{2017Pinilla}  hints that there could be substructures present within these disks that remain undetected, allowing for higher dust disk masses to be retained. This highlights the importance of higher angular resolution observations of the sample.

In comparison with the dust masses obtained from the multiwavelength analysis in Pulgarés et al (in prep) our dust masses are even more underestimated. The low dust masses in our models are likely to stem from our assumed initial conditions. We initialize each simulation with a dust-to-gas ration of 1\% based on the \textit{current} gas mass of the disk, as inferred from the AGE-PRO modeling by \cite{Trapman_2025_gas}. However, if the true initial gas mass was much larger, as expected if disks lose gas over time through winds or accretion, then the initial dust reservoir will also have been correspondingly larger. The underprediction in our model therefore reflects the uncertainty of the \textit{initial} gas mass.

\begin{figure*}
    \centering
    \includegraphics[width=0.96\textwidth]{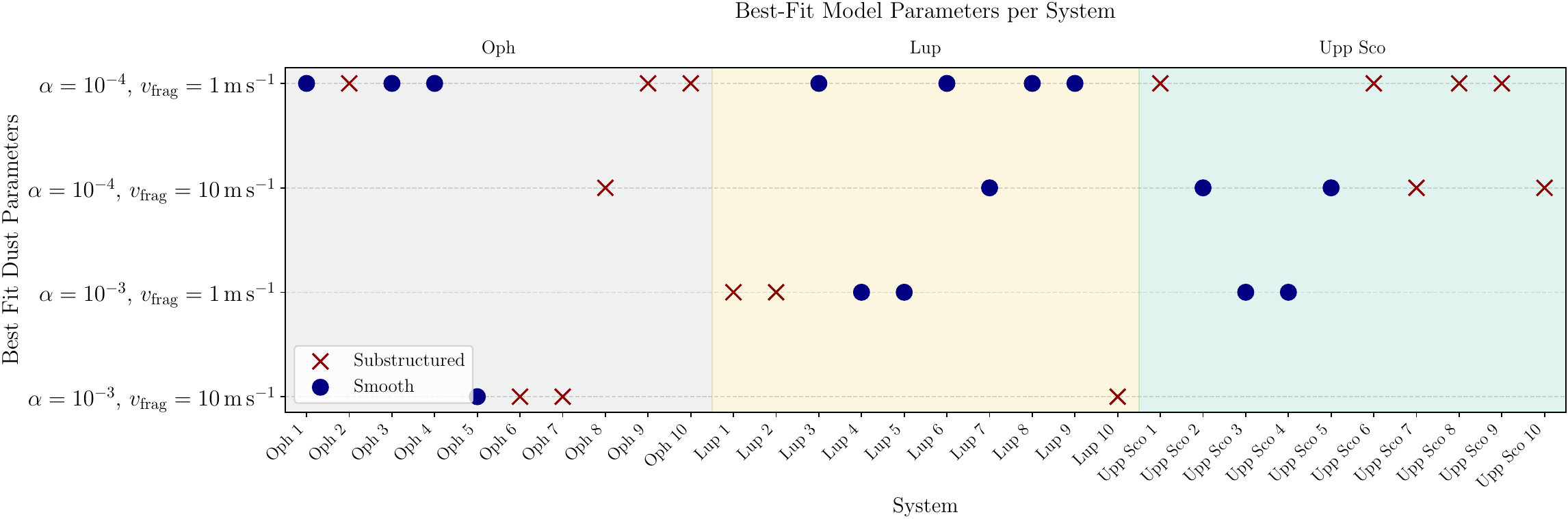}
    \caption{Summary of the best-fitting dust evolution model parameters for each disk based on $\chi^2$ minimization of the radial intensity profiles modeled disks compared with that of FRANK radial profile. Markers distinguish between substructured (×) and smooth (o) disk models, while the y-axis is the corresponding parameter set. Systems are grouped by star-forming region, indicated by grey (Ophiuchus), golden (Lupus), and teal (Upper Scorpius) background shading.
    }
    \label{fig:best_fit}
\end{figure*}

\begin{table}
\centering
\caption{Observed median dust disk masses for smooth and substructured disks in each star-forming region of the AGE-PRO sample. Substructured disks exhibit systematically higher dust masses than smooth disks in Lupus and Upper Scorpius.}
\label{tab:obs_mdust_sub_vs_smooth}
\begin{tabular}{lcc}
\hline
Region & Smooth Median [$M_\oplus$] & Substructured Median [$M_\oplus$] \\
\hline
Ophiuchus & 7.5  & 16.5 \\
Lupus     & 3.0  & 50.0 \\
Upper Sco & 0.7  & 8.4  \\
\hline
\end{tabular}
\end{table}

We quantify the difference in dust disk mass observations of the smooth and substructured disks by computing median dust masses for each region (Table~\ref{tab:obs_mdust_sub_vs_smooth}). We find that in observations, substructured disks consistently exhibit higher dust masses than smooth disks in Lupus and Upper Sco, with differences of more than an order of magnitude for both regions. This contrast is much weaker in the younger Ophichius region. This trend supports the scenario in which pressure traps associated with substructures slow radial drift and enable long term dust mass retention, particularly at later evolutionary stages.

\subsection{\texorpdfstring{Finding the best $\alpha$ and $v_{\rm{frag}}$}{Finding the best alpha and vfrag}}
In order to further investigate the optimal combinations of $\alpha$ and $v_{\rm{frag}}$, we compare the radial profiles of our models with the FRANK profiles. To do this, we first generate synthetic face-on Band 6 images of each model. These images are then convolved with a circular Gaussian that has a FWHM that matches the effective resolution of the FRANK reconstruction for the corresponding disk. Then we extract normalized radial profiles from the convolved images and compare them directly to the normalised FRANK profile.

 We compute the reduced $\chi^2$ from the difference between the two profiles and the $\alpha$ and $v_{\rm{frag}}$ combination with the lowest $\chi^2$ is selected as the best fit for that disk. Our best fit combinations of $\alpha$ and $v_{\rm{frag}}$ are presented in Fig.\ref{fig:best_fit}. We find that roughly half of the disk favor $\alpha = 10^{-4}$ and $v_{\rm{frag}} =1\,\mathrm{m\,s^{-1}} $. We find no correlation between the best fit combinations of $\alpha$ and $v_{\rm{frag}}$ and whether the disk is smooth or substructured. We also check the best fit forcing $v_{\rm{frag}}$ to be one of the two assumed values, in order to find a unique $\alpha$, however we find that when we do that, the best fit are evenly spread among the two values of $\alpha$.

\subsection{Pebble Flux and Cold Water Delivery} \label{sect:results_pebble_flux}

\begin{figure*}
    \centering
    \includegraphics[width=0.96\textwidth]
    {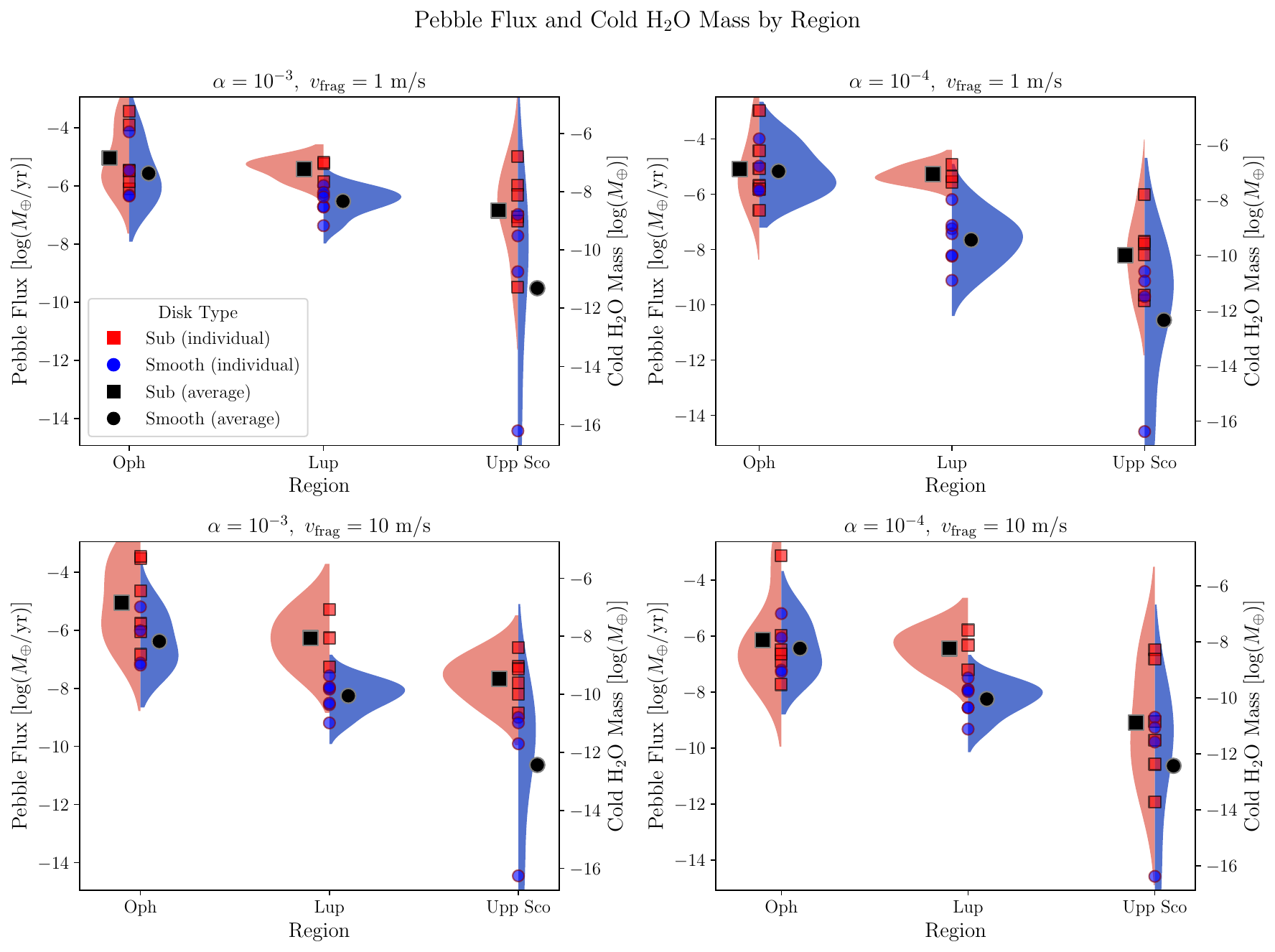}
    \caption{Pebble flux (left-hand y-axis) and corresponding cold water mass (right-hand y-axis) delivered to the inner disk (within $\sim$1 AU). Both vertical axes are shown on a logarithmic scale. Results are grouped by star-forming region (Ophiuchus, Lupus, and Upper Sco) and by the four combinations of $\alpha$ and $v_{\mathrm{frag}}$. The estimated cold water mass is proportional to the pebble flux derived from the dust evolution models. Each violin plot shows the distribution of predicted cold water masses for substructured disks (red, left half) and smooth disks (blue, right half). The black square indicates the median value for the substructured subset, while the black circle shows the median for the smooth disks. The overall trend shows reduced cold water delivery in older regions, consistent with declining dust mass and pebble drift efficiency over time.
}
    \label{fig:ColdWater_pebbleflux}
\end{figure*}

Figure~\ref{fig:ColdWater_pebbleflux} presents a comparative overview of pebble flux and the associated cold H$_2$O mass reaching the inner disk (within $\sim$5 AU). As the two quantities are linearly related in our model framework (Equation~\ref{eq:watermass}), we show pebble flux on the left hand side axis and the cold H$_2$O mass on the right hand side. The plot is organized by model configuration and grouped by region.

Our results show a clear decline in both pebble flux and cold water mass with increasing disk age. Disks in Ophiuchus exhibit systematically higher delivery rates than those in Lupus and Upper Scorpius. This age-driven trend reflects the diminishing dust reservoir and the progressive inward drift and depletion of mm-sized grains over time.

Figure~\ref{fig:ColdWater_pebbleflux} also shows that at fixed $\alpha$ and $v_{\rm{frag}}$ substructured disks will typically exhibit higher pebble fluxes at the snowline (and hence also higher inferred cold water masses) than disks we modeled as smooth. At the older ages represented by Lupus and Upper Sco, this difference can reach $\sim 2$--$3$ orders of magnitude. Within our framework, this separation reflects two coupled effects: (i) the substructured subset is, on average, more massive in our sample (as shown in Table~\ref{tab:obs_mdust_sub_vs_smooth}), and (ii) leaky traps can prolong the survival of a drifting grain reservoir by continuously replenishing small grains through fragmentation, thereby sustaining the transport of material inward at later times. Even when normalizing to the initial pebble flux of each disk (see Appendix Figure~\ref{fig:pebbleflux_initial}), substructured models maintain higher fluxes at later times (Lupus and Upper Sco). This indicates that disk morphology itself (i.e. leaky traps) will contribute to the persistence of late-time pebble flux, but time evolution has the strongest effect on depleting the water delivery to the inner disk. For the case of $\alpha=10^{-3}$ and $v_{\mathrm{frag}} = 10\,\mathrm{m\,s^{-1}}$, radial drift is already efficient at the age of Ophiuchus leading to a higher pebble flux for substructured disks even after normalizing to initial pebble flux. Our results are consistent with dust evolution simulations of \cite{Pinilla_2026}, who compared pebble fluxes of disks with and without a trap. They demonstrated that disks without traps exhibit a steady decline in flux as the outer reservoir is depleted, while disks with traps converge to toward a sustained, nearly constant pebble flux at later times. If there is a relationship between pebble flux and cold water, this implies that the observable difference between smooth and substructured disks is a time-dependent signal.

\cite{banzatti2023, banzatti2025} reported a correlation between the luminosity of cold water lines and the dust disk size as measured by ALMA, finding that larger dust disks tend to exhibit lower cold water luminosities. This has been interpreted as evidence that large disks retain ice-bearing pebbles in pressure bumps, whereas in small, smooth disks pebbles drift efficiently inward and transport ices to the inner disk. However, not all observational surveys confirm this trend, for example, recent MINDS result \citep{2025Temmink}, did not find evidence for such a correlation.
We investigate whether there is any correlation between the cold water masses shown in Fig.~\ref{fig:ColdWater_pebbleflux} and $R_{90\%}$ in our models. However, we do not recover the trend suggested by \cite{banzatti2023, banzatti2025}. 

\begin{figure*}
    \centering
    \includegraphics[width=0.96\textwidth]{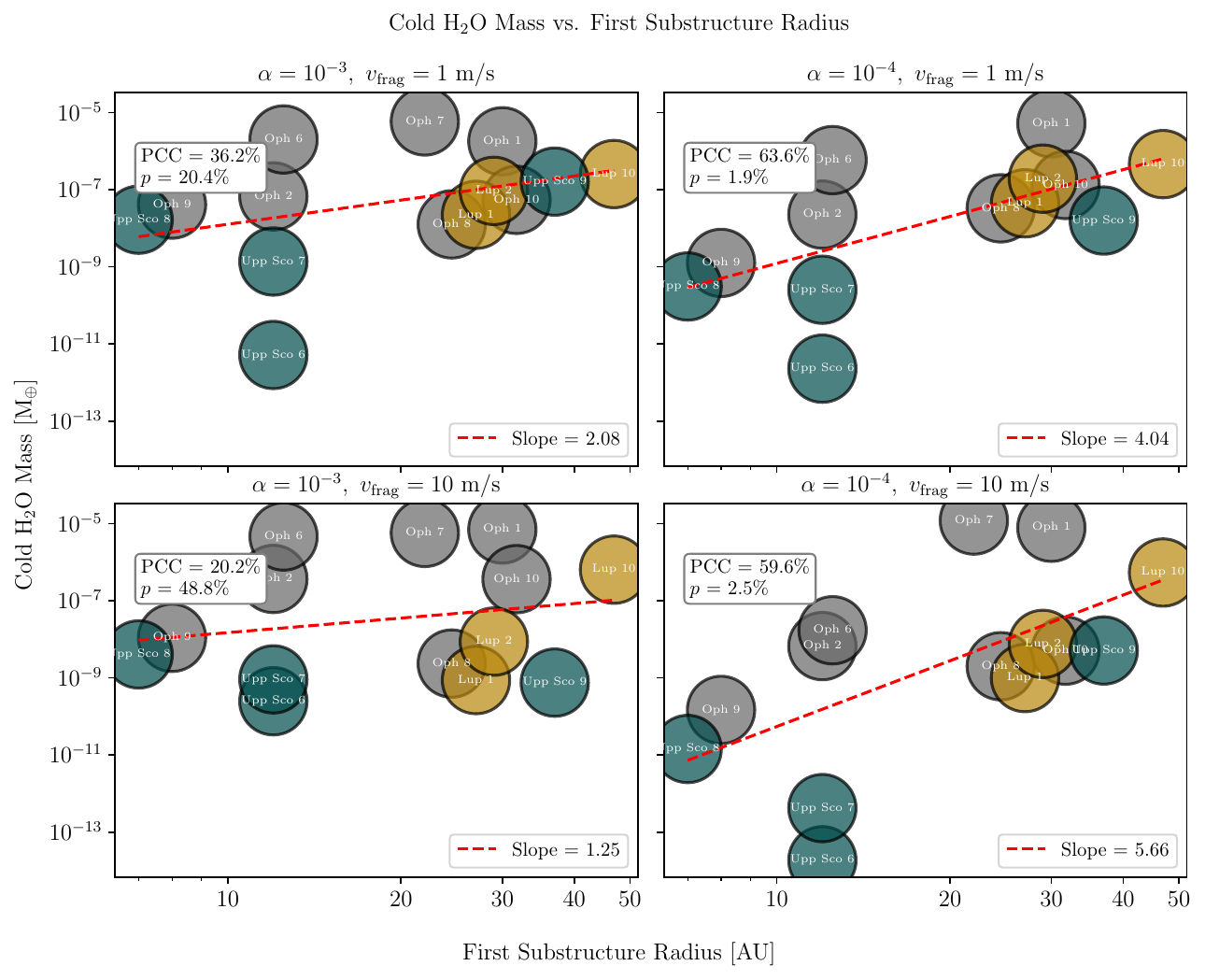}
    \caption{We present the cold water masses of the substructured disk sample as a function of the location of the innermost substructure, for the four combinations of $\alpha$ and $v_{\rm{frag}}$. The dashed line in each subplot indicates a linear regression fit to the data, with the associated slope, p-value, and Pearson correlation coefficient (PCC) reported in the legend. }
    \label{fig:Coldwater_substructure_rad}
\end{figure*}

\cite{kalyaan2023, Mah_2024, 2024Easterwood} suggested a correlation between the cold water abundance and the location of the innermost bump or gap, provided that it lies outside the snowline. The closer the trap is to the snowline, the more efficiently pebbles are retained in the outer disk. In our models (see Fig. \ref{fig:Coldwater_substructure_rad}) we find a positive correlation between the location of the first substructure and amount the of cold water in the inner disk, such that disks with a more distant first substructure contain more water. Measuring the slope of this correlation observationally could help to discriminate between models, as we find it varies for different values of $\alpha$ and $v_{\rm{frag}}$. The Pearson correlation coefficient is the highest for the models with $\alpha=10^{-4}$. Although the correlation is strongest for $\alpha=10^{-4}$, such low turbulence levels are generally insufficient to reproduce the observed accretion rates. This could point to the need for additional mechanisms, such as layered accretion or magnetically driven winds, as discussed before.

We also note that the resolution of the AGE-PRO sample from the FRANK fitting is around 10-50 AU, which means that we may still miss substructures inside, which can affect the results of our dust evolution models. 

\section{Discussion} \label{sect:discussion}

\subsection{Observational trends versus models} \label{sect:discussion_1}

In this work, we compare the trends obtained from observations of the AGE-PRO sample \citep{zhang2025} to dust evolution models to understand if the overall behavior of the $R_{90\%}$, spectral index ($\alpha_{\rm{mm}}$), dust disk mass, and cold water abundance in the inner disk may favor one of the combinations of $\alpha$ and $v_{\rm{frag}}$.  We also compare these results with the multiwavelength  analysis of the AGE-PRO sample from Pulgarés et al. (in prep).

For each of the combinations of $\alpha$ and $v_{\rm{frag}}$, there is not a very significant variation of $R_{90\%}$ across the regions (see red violin plots from Figure~\ref{fig:Violin_per_bands}), in agreement with the results from \cite{zhang2025}. Consistent with the findings of \cite{zhang2025}, there seems to be a slight decrease in $R_{90\%}$ (Band 6) from Oph to Lupus. 
Across the full AGE-PRO sample, our models show a weak dependence on wavelength, and particularly in the case of Lupus there is almost no variation with wavelength (see Figure ~\ref{fig:Violin_per_bands}). The observational analysis of Pulgarés et al. (in prep) finds that $R_{90\%}$ is wavelength independent. While we find that there is a weak dependence on wavelength for the full sample we note that this is mainly driven by smooth disks or disks with weaker pressure bumps (see Oph 2 Figure~\ref{fig:spectralR90}), in contrast disks with stronger bumps (see Upp Sco 1 Figure~\ref{fig:spectralR90}), exhibit little or no wavelength dependence because grains of all sizes are trapped at similar radii.

 \cite{2025_Vioque} suggested that the reason why $R_{90\%}$ does not change with age is because of the presence of traps. Our results suggest that even when only half of the sample has substructures included in the models, $R_{90\%}$ is approximately constant with age and wavelength, and independent of the assumed $\alpha$ and $v_{\rm{frag}}$. This means that our models reproduce fairly well the observed trends for $R_{90\%}$, but the evolution of $R_{90\%}$ or its change across different wavelengths do not contribute to constrain $\alpha$ or $v_{\rm{frag}}$.

For the spectral index, Ophiuchus has the lowest value of $\alpha_{\rm{mm}}$  compared to Lupus and Upper Sco for any combination of $\alpha$ and $v_{\rm{frag}}$ (see Fig.~\ref{fig:spectralB}. This is because at the age assumed for Ophiuchus (0.5\,Myr), most of the dust is still in the disk, and in the outer disk the dust particles are still growing to sizes that will drift efficiently at later times. This difference between Ophiuchus and Lupus/Upper Sco is not seen in the observations. In addition, our spectral indices are higher than observed, either because our radiative transfer models do not include contributions from no thermal-emission or because our models do not include all the traps that these disks may actually have. \cite{pinilla2021} show that by increasing the number of pressure bumps in a disk, the spectral index is expected to decrease. As the behavior of $\alpha_{\rm{mm}}$  is similar for any combination of $\alpha$ and $v_{\rm{frag}}$ in our models, $\alpha_{\rm{mm}}$  does not contribute to constrain these values.  

Comparing the model predictions with the observations, overall trends in dust disk mass favor $\alpha=10^{-3}$ and $v_{\mathrm{frag}} = 1\,\mathrm{m\,s^{-1}}$, as this is the only set of models where the dust disk mass decreases across regions, which is consistent with the measurements from AGE-PRO (see Fig. ~\ref{fig:Mdust_violin}). In the other three sets of models, the dust disk mass between Lupus and Upper Sco is similar. However, the set of models with $\alpha=10^{-3}$ and $v_{\mathrm{frag}} = 1\,\mathrm{m\,s^{-1}}$ is one of the least favored in Fig.~\ref{fig:best_fit}, when comparing radial profiles. We note, however, that our dust disk masses are underestimated, especially when comparing with the multi-wavelength analysis from Pulgarés et al.(in prep), which can be due to our initial condition assumptions, and the lack of substructures in the smooth disks, as discussed in Sect.~\ref{sect:methods}. We highlight the importance to characterize potential substructures in the smooth disks in future observations.

Finally, although the general trend of the pebble flux to the inner disk (and the cold water abundance) is similar for any combination of $\alpha$ and $v_{\rm{frag}}$ that is assumed in the models, we note the absolute values can vary by up to 2 orders of magnitude between configurations (see Fig. ~\ref{fig:ColdWater_pebbleflux}). Nevertheless, a potential correlation of the cold water abundance with the location of the inner trap in the AGE-PRO sample   may help constrain $\alpha$ and $v_{\rm{frag}}$. This relation is expected to be the strongest for low values of $\alpha$ (see Fig. ~\ref{fig:Coldwater_substructure_rad}), which is consistent with our radial brightness profile fitting, which favors low $\alpha$ for the majority of the disks. 

\subsection{Implications for Pebble Accretion}

To assess whether the modeled pebble fluxes are sufficient to support planet formation, we compare them to thresholds derived from pebble accretion theory. In our models, we estimate the snowline pebble mass flux through the disk and examine its distribution as a function of time. Although we find that substructured disks tend to maintain higher pebble fluxes into the inner few AU compared to smooth disks, the absolute values (typically $\dot{M}{\rm peb} \sim 10^{-9} - 10^{-7}~M_\oplus~{\rm yr}^{-1}$) appear low relative to those required to form sub-Neptune mass planets. For example, \citet{kalyaan2023} and \citet{2014Lambrechts} show that building a $\sim$2–5 $M_\oplus$ core within typical disk lifetimes (1–5 Myr) requires sustained pebble fluxes of at least $10^{-7}M_\oplus{\rm yr}^{-1}$ under efficient accretion conditions. Integrating our model fluxes over time yields cumulative pebble masses of $\sim$2\,$M_\oplus$ in the median, but with a mean of $\sim$50\,$M_\oplus$, and some models reaching up to $\sim$1000\,$M_\oplus$. This large spread suggests that while certain disk structures can enable significant inward transport of solids, many models might still fall short of delivering enough material to form sub-Neptune planets unless accretion is highly efficient. These results may motivate further coupling of pebble flux models with planet formation efficiency estimates to assess whether the observed water-rich inner disks can be explained by the available solid mass budget.

As mentioned in Section ~\ref{sect:results_dust_disk_mass}, our potentially low pebble fluxes may result from the use of \textit{current} gas mass to initialize the 1\% dust-to-gas ratio in each model. If the true initial gas masses were substantially larger, then the initial dust reservoirs would also have been larger, leading to more solid mass available for pebble drift. The error between initial and current gas mass is more significant for older systems. Future models that incorporate time-varying gas masses or that draw initial conditions based on population synthesis will be necessary to give more accurate estimates of the pebble fluxes and water-delivery efficiencies.

As discussed in ~\ref{sect:results_pebble_flux}, we interpret the systematically higher pebble fluxes found for substructured disks at late evolutionary stages as a consequence of leaky dust traps being able to sustain the flux into the inner disk for a longer time period.  This is also consistent with the modeling of \cite{Pinilla_2026}. In our models, the substructured subset of disks are initially more massive, hence brighter and easier to resolve substructures, while smooth disks are fainter and they also appear smaller at the current angular resolution. Therefore structured disks exhibit higher pebble fluxes at early times, before radial drift significantly depletes the outer disk. At later ages, corresponding to Lupus and Upper Scorpius regions, the presence of pressure traps allows substructured disks to sustain a long-lasting delivery of pebble flux into the inner disk. This effect arises since dust trapped in pressure bumps will undergo continuous fragmentation, replenishing a population of small grains that can diffuse inward with the gas when the trap is partially leaky. In contrast, smooth disks lacking such traps experience much more efficient radial drift that rapidly depletes the outer dust reservoir, leading to a strong decline in pebble flux over time. This results in older substructured disks retaining higher pebble fluxes compared with smooth disks of comparable age. We therefore interpret the enhanced pebble fluxes observed in older substructured disks as a direct consequence of long-lived dust reservoirs enabled by pressure traps, whereas the differences seen in younger disks mainly reflect the initial disk mass distribution. This implies that any observational link between inner-disk water abundances and disk substructures is expected to be the most apparent when comparing disks across a broad range of evolutionary stages, as concluded by  \cite{Pinilla_2026}. 

\subsection{\texorpdfstring{Universality of $\alpha-$viscosity and $v_{\rm{frag}}$}{Universality of alpha-viscosity and vfrag}} 

To investigate whether specific combinations of $\alpha$ and $v_{\rm{frag}}$ are generally favored in reproducing observed disk properties, we compare the radial profile predictions of our four combinations of $\alpha$ and $v_{\rm{frag}}$ in our models for each disk against observations. We find that the combination of $\alpha =10^{-4}$ and $v_{\rm{frag}} =1\rm{m s}^{-1}$, provides the best fit to approximately half of the observed radial brightness profiles, derived from the FRANK modeling. \boldred{Interestingly, a similar preference for $\alpha =10^{-4}$ and $v_{\rm{frag}} =1\rm{m s}^{-1}$ was also reported by \citet{2024Jiang} based on multi-wavelength constraints on grain sizes.}

Although such low $\alpha$ values are favored by dust evolution models, they are challenging to reconcile with observed accretion rates \citep{Tabone_2025}. This tension has motivated alternative accretion drivers such as magnetized winds, which are increasingly supported by observational evidence \citep{2020Pascucci}.

However, when we examine the broader trends across the full sample  as discussed in Sect. ~\ref{sect:results}, ~\ref{sect:discussion_1}, we find that disk properties, such as $M_{\mathrm{dust}}$, spectral indices, and $R_{90\%}$ do not consistently favor a single $\alpha$ and $v_{\rm{frag}}$ combination. In particular, the configuration that best reproduces the dust masses does not necessarily match the best fit for $R_{90\%}$ or $\alpha_{\rm{mm}}$. This indicates that no single combination of $\alpha$ and $v_{\rm{frag}}$ performs the best across all diagnostics.

Moreover, we do not observe a clear correlation between age of the disk and the preferred $\alpha$, $v_{\rm{frag}}$ combination. Younger disks (Ophiuchus) do not uniformly favor different parameters when compared to older disks (Lupus, Upper Sco). This lack of age dependence may indicate, within the parameter space we explored, that disk evolution alone does not select for a specific turbulence strength or fragmentation limit. 

Finally, we also acknowledge that $\alpha$ and $v_{\rm{frag}}$ may vary spatially and temporally within individual disks. For example, accretion or dead zones may cause local variations in turbulence, while $v_{\rm{frag}}$ may vary with an evolving dust composition or with icy grains. Our results suggest, however, that the diversity in disk structure and evolution seems to cover a range of possible physical conditions rather than have a universal prescription.

\subsection{Future perspectives}

For the \texttt{DustPy} models, we assume a fixed gas surface density rather than evolving gas and dust simultaneously. This was a matter of simplicity as it is still unclear what drives the evolution of the gas, and in Upper Sco, the disks can also be subject to external photoevaporation \citep{2025Anania}. The initial gas disk mass in each disk would need to be higher in the models than the value obtained by \citep{Trapman_2025_gas}, with the degree to which the gas mass is underestimated being dependent on what is the main driver of the gas evolution. Given these uncertainties, we kept the gas disk mass constant over time to ensure that the dust-gas coupling remains consistent for the grain populations probed at different wavelengths. We highlight that performing gas and dust evolution models simultaneously are key to better constrain  $\alpha$  and $v_{\rm{frag}}$.

From an observational perspective, higher-resolution observations capable of revealing potential hidden substructures would help determine whether the observed trends remain, and consequently whether our conclusions about $\alpha$, $v_{\rm{frag}}$, and pebble fluxes remain valid when compared with models. Other diagnostics that can provide constraints on turbulence can also benefit from the understanding of dust evolution and the key parameters that play a role. For example, potential high-resolution observations of the highly inclined disks in the AGE-PRO sample, could allow us to measure the level of vertical turbulence by fitting their substructures as recently done for other disks \citep{Villenave2025}.
We acknowledge, however, that the level of turbulence inferred from vertical settling may differ from that of the radial angular momentum transport. There is observational evidence supporting wind-driven accretion could play an important role in disk evolution (e.g., \cite{2023Pascucci}; \cite{2025Pascucci}), offering an alternative or complementary mechanism to turbulence.

\section{Conclusions} \label{sect:conclusion}

In this work we modeled 30 disks in dust evolution and explored the impact of different combinations of $\alpha$  and $v_{\rm{frag}}$, with the potential of setting constraints on them. In addition, we also give predictions of inner pebble flux or cold water abundance for future comparisons with JWST observations. Our main conclusions are:

\begin{enumerate}
    \item The observed evolutionary trend for $R_{90\%}$, spectral index, and dust disk mass do not favor any combination of $\alpha$  and $v_{\rm{frag}}$. In most cases, any combination $\alpha$  and $v_{\rm{frag}}$ can explain the observations. 
    \item By comparing the radial intensity profile of each disk for each combination of $\alpha$  and $v_{\rm{frag}}$, the majority of the disks would be better explained with values of $\alpha=10^{-4}$. 
    \item Across all $\alpha$  and $v_{\rm{frag}}$ combinations, and for both smooth and substructured disks, time evolution is the dominant driver of inner pebble flux decline. Early-time fluxes are largely independent of whether the disk has substructures or not, with substructures only becoming a distinguishable effect at later ages. Testing the pebble flux-water abundance connection therefore requires samples spanning a wide range of ages.
    \item We predict a correlation between the cold water abundance and the location of the innermost trap in the disks, which slope and strength varies for each combination of $\alpha$  and $v_{\rm{frag}}$. The correlation is expected to be stronger with $\alpha=10^{-4}$.

\end{enumerate} 

We highlight the importance of future work on modeling gas and dust evolution simultaneously, including different physical processes such as MHD winds and external photoevaporation. Future high angular resolution observations for the potential detection of hidden structures could further help to constrain key parameters of dust evolution processes.

\section*{Acknowledgements}

LL would like to thank the Science and Technology Facilities Council (STFC) for funding support through a PhD studentship.

PP and AS acknowledge funding from the UK Research and Innovation (UKRI) under the UK government’s Horizon Europe funding guarantee from ERC (under grant agreement No 101076489). 

\boldred{L.P. acknowledges support by the ANID BASAL project FB210003 and ANID FONDECYT Regular 1221442.}

C.A.G. acknowledges support from Comité Mixto ESO-Gobierno de Chile 2023, under grant 072-2023. 

GR acknowledges funding from the Fondazione Cariplo, grant no. 2022-1217, and the European Research Council (ERC) under the European Union’s Horizon Europe Research \& Innovation Programme under grant agreement no. 101039651 (DiscEvol). Views and opinions expressed are however those of the author(s) only, and do not necessarily reflect those of the European Union or the European Research Council Executive Agency. Neither the European Union nor the granting authority can be held responsible for them.

JM acknowledges support from ANID – Millennium Science Initiative Program – Center Code NCN2024\_001 and from
FONDECYT de Postdoctorado 2024 \#3240612 .

L.A.C acknowledges support from the Millennium Nucleus on Young Exoplanets and their Moons (YEMS), ANID - Center Code NCN2024\_001 and from the FONDECYT grant \#1241056.

\section*{Data Availability}

The data underlying this article are available from the corresponding author upon request.

\bibliographystyle{mnras}
\bibliography{example}

\appendix

\section{The effect of substructures on pebble flux}

\begin{figure*}
    \centering
    \includegraphics[width=0.96\textwidth]
    {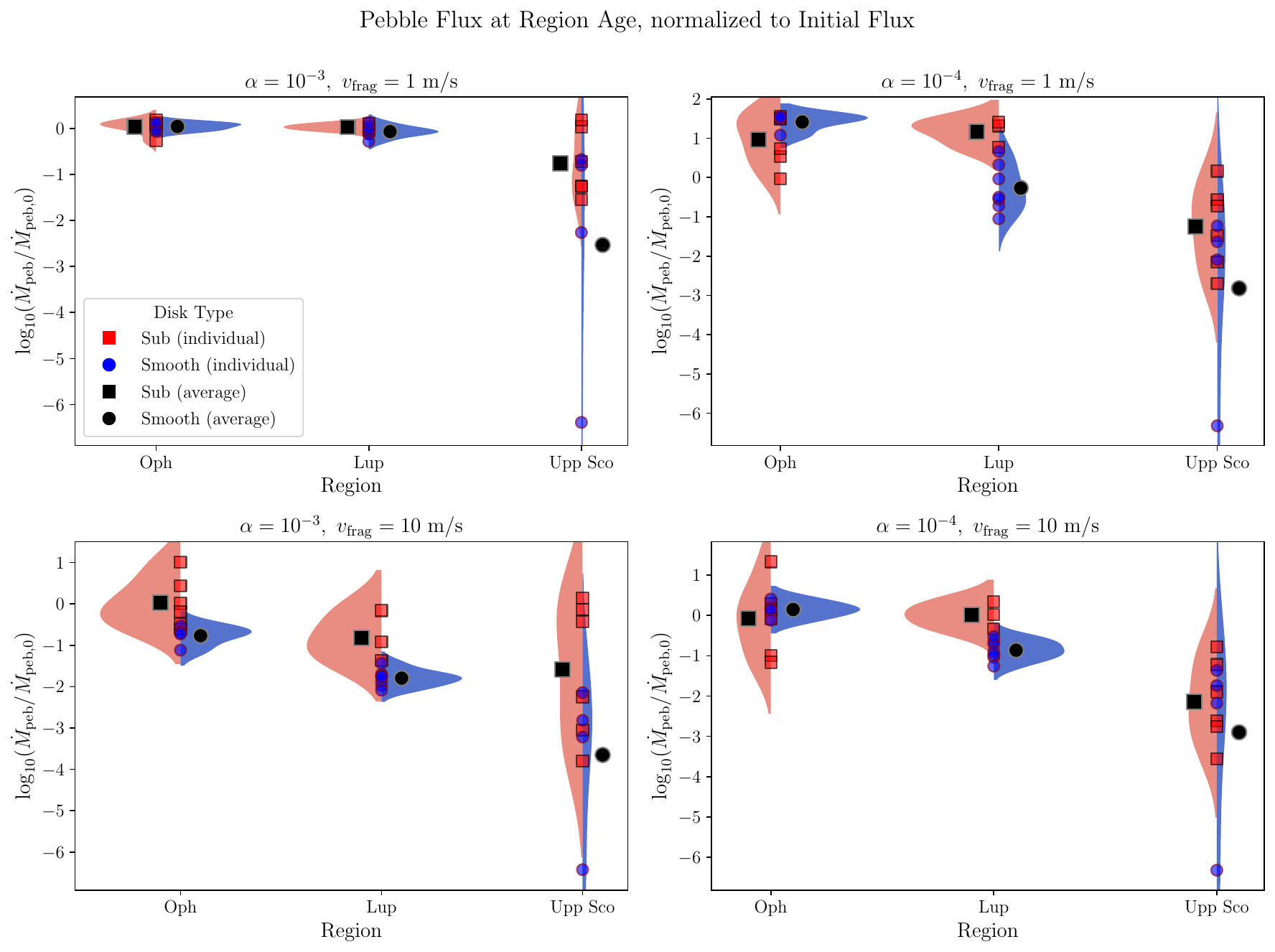}
    \caption{Pebble flux at snowline (within $\sim$1 AU) normalized to initial pebble flux, shown for each star-forming region (Ophiuchus, Lupus, and Upper Sco) and for the four combinations of $\alpha$ and $v_{\mathrm{frag}}$. The vertical axis therefore represents a dimensionless ratio, and it is shown on a logarithmic scale. Each violin plot shows the distribution of predicted normalized pebble fluxes for substructured disks (red, left half) and smooth disks (blue, right half). The black square indicates the median value for the substructured subset, while the black circle shows the median for the smooth disks. Normalizing by the initial pebble flux reduces the separation between smooth and substructured disks at young ages. In most parameter combinations, there is little difference at Ophiuchus ages, while a clearer separation emerges at later times (Lupus and Upper Sco), where substructured disks tend to retain higher relative fluxes.}
    \label{fig:pebbleflux_initial}
\end{figure*}

\begin{figure*}
    \centering
    \includegraphics[width=0.96\textwidth]
    {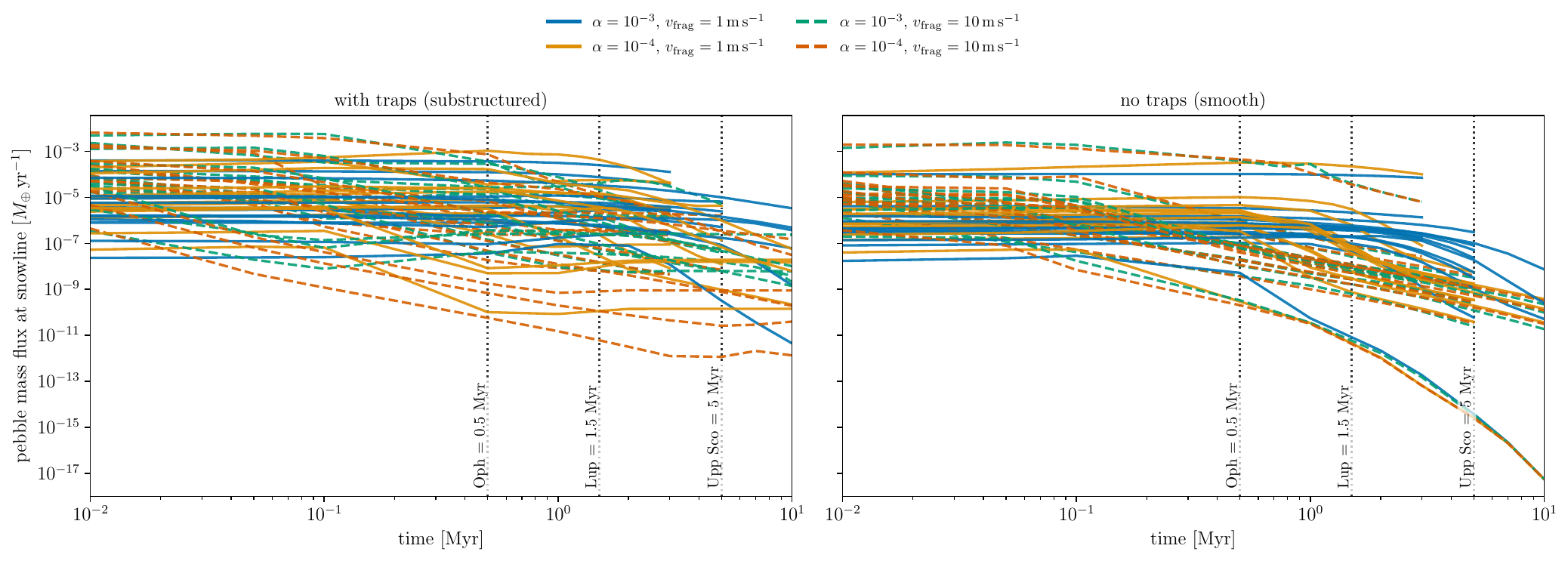}
    \caption{Pebble flux at snowline as a function of time. The LHS plot consists of the disks treated as substructured, while the RHS consists of the disks considered smooth.  The legend indicates the four combinations of $\alpha$ and $v_{\mathrm{frag}}$, and each disk is labeled in the figure. We have also indicated the ages of the regions. There is a large spread in pebble flux for the disk when varying $\alpha$ and $v_{\mathrm{frag}}$. Our findings are similar to that of the trends observed by \citep{Pinilla_2026}. The overall trend shows reduced cold water delivery in older regions, consistent with declining dust mass and pebble drift efficiency over time. While the pebble flux in the smooth disks decrease rapidly the flux in the substructured disks persists for longer.}
    \label{fig:pebbleflux_timeseries}
\end{figure*}

%%%%%%%%%%%%%%%%%%%%%%%%%%%%%%%%%%%%%%%%%%%%%%%%%%

\bsp	
\label{lastpage}
\end{document}